\def\eop{E^{\rm obs}_{\rm peak}}

\def\ep{E_{\rm peak}}

\documentclass[12pt,preprint]{aastex}

\shorttitle{BAT catalog}
\shortauthors{Sakamoto et al.}

\begin{document}


\title{The First {\it Swift} BAT Gamma-Ray Burst Catalog}


\author{
T. Sakamoto\altaffilmark{1,2}, 
S. D. Barthelmy\altaffilmark{1}, 
L. Barbier\altaffilmark{1}, 
J. R. Cummings\altaffilmark{3,10}, 
E. E. Fenimore\altaffilmark{4},
N. Gehrels\altaffilmark{1}, 
D. Hullinger\altaffilmark{8}, 
H. A. Krimm\altaffilmark{6,10}, 
C. B. Markwardt\altaffilmark{5,10},
D. M. Palmer\altaffilmark{4},
A. M. Parsons\altaffilmark{1},
G. Sato\altaffilmark{1,7},
M. Stamatikos\altaffilmark{1,2}, 
J. Tueller\altaffilmark{1},
T. N. Ukwatta\altaffilmark{1,9},
B. Zhang\altaffilmark{11}
}

\altaffiltext{1}{NASA Goddard Space Flight Center, Greenbelt, MD 20771}
\altaffiltext{2}{Oak Ridge Associated Universities, P.O. Box 117, 
 Oak Ridge, Tennessee 37831-0117}
\altaffiltext{3}{Joint Center for Astrophysics, University of Maryland, 
	Baltimore County, 1000 Hilltop Circle, Baltimore, MD 21250}
\altaffiltext{4}{Los Alamos National Laboratory, P.O. Box 1663, Los
Alamos, NM, 87545}
\altaffiltext{5}{Department of Astronomy, University of Maryland, 
	College Park, MD 20742}
\altaffiltext{6}{Universities Space Research Association, 10211 Wincopin 
	Circle, Suite 500, Columbia, MD 21044-3432} 
\altaffiltext{7}{Institute of Space and Astronautical Science, 
JAXA, Kanagawa 229-8510, Japan}
\altaffiltext{8}{Moxtek, Inc., 452 West 1260 North, Orem, UT  84057}
\altaffiltext{9}{Center for Nuclear Studies, Department of Physics, 
	The George Washington University, Washington, D.C. 20052}
\altaffiltext{10}{Center for Research and Exploration in Space Science 
and Technology (CRESST), NASA Goddard Space Flight Center, Greenbelt, MD 
20771}
\altaffiltext{11}{Department of Physics and Astronomy, University of Nevada, 
Las Vegas, NV 89154}



\begin{abstract}
We present the first {\it Swift} Burst Alert Telescope (BAT)
catalog of gamma-ray bursts (GRBs), which contains bursts detected by the BAT
between 2004 December 19 and 2007 June 16.  This catalog (hereafter BAT1 
catalog)
contains burst trigger time, location, 90\% error radius, duration, 
fluence, peak
flux, and time averaged spectral parameters for each of 237 GRBs, as measured
by the BAT.  The BAT-determined position reported here is within 1.75$^{\prime}$ of
the Swift X-ray Telescope (XRT)-determined position for 90\% of these GRBs.
The BAT T$_{90}$ and T$_{50}$ durations peak at 80 and 20 seconds, respectively.  
From the fluence-fluence correlation, we conclude that about 60\% of the
observed peak energies, $\eop$, of BAT GRBs could be less than 
100 keV.  We confirm that GRB fluence to hardness and GRB peak flux to hardness 
are correlated for BAT bursts in analogous ways to previous missions' results.  
The correlation between the photon index in a simple power-law model 
and $\eop$ is also confirmed.  We also report the 
current status for the on-orbit BAT calibrations based on observations of the 
Crab Nebula.  

\end{abstract}





\section{Introduction}

The {\it Swift} mission \citep{gehrels2004} has revolutionized our 
understanding of gamma-ray bursts (GRBs).  Because of the sophisticated 
on-board localization capability of the {\it Swift} Burst Alert Telescope 
(BAT; \citet{barthelmy}) and the fast spacecraft pointing of {\it Swift}, 
more than 90\% (30\%) of {\it Swift} GRBs have an X-ray (optical) 
afterglow observation from the {\it Swift} X-Ray Telescope (XRT;
\citet{burrows}) or the {\it Swift} UV/Optical Telescope (UVOT; \citet{roming}) 
within a few hundred seconds after the trigger.  ${\it Swift}$ opens 
a new opportunity to study the host galaxies and the distance scale 
of the mysterious short duration GRBs.\footnote{The short burst GRB 050709, 
which observed and promptly localized by {\it HETE-2} is also a good 
example of a short GRB observation \citep{villasenor2005}.} 
\citep[e.g.,][]{gehrels2005,barthelmy2005b}  {\it Swift} allows us to use 
GRBs as a tool for investigation of the early universe 
(cf. detection of GRB 050904 at a redshift of 6.29; \citet{cusumano2006}).  
{\it Swift} found the fourth GRB, GRB 060218, which is securely associated with 
a supernova \citep{campana2006}.  
Furthermore, {\it Swift} is providing X-ray 
afterglow data \citep[e.g.,][]{zhang2006} which gives us insight into details 
of the fireball model and to the nature of the central engine.

Here we present the first BAT GRB catalog including 237 GRBs detected 
by BAT from 2004 December 19 to 2007 June 16.  In \S 2, we describe 
the BAT instrument.  In \S 3, we show the current 
status of the on-orbit calibration of the BAT based on the Crab observations.  
In \S 4, we describe the analysis methods for the catalog.  In \S 5, we 
describe the content of the tables of the catalog and show the results of the 
prompt emission properties of the BAT GRBs based on the catalog.  Our 
conclusions are summarized in \S 6.  All quoted errors in this work 
are at the 90\% confidence level.  

\section{Instrumentation}

The BAT is a highly sensitive, large field of view (FOV) (1.4 sr for $>$50\%
coded FOV and 2.2 sr for $>$10\% coded FOV), coded-aperture 
telescope, which detects and localizes GRBs in real time.  The fast and accurate 
BAT GRB positions with 1-3 arc-minute error radii are the key to 
autonomously slewing the 
spacecraft to point the XRT and the UVOT.  The BAT GRB position, light
curves, and the detector plane image (BAT scaled map) are transmitted through 
TDRSS to the ground within 20--200 s after the burst trigger.  
The BAT detector plane is 
composed of 32,768 pieces of CdZnTe (CZT: $4 \times 4 \times 2$ mm), and the 
coded-aperture mask is composed of $\sim$ 52,000 lead tiles 
($5 \times 5 \times 1$ mm) with a 1-m separation between mask and 
detector plane.  The energy range is 15-150 keV for imaging or mask
weighting\footnote{Mask weighting is a
background subtraction technique based on the modulation resulting from
the coded mask.} with a non-coded response up to 350 keV.  

The CZT pixels are biased to $-200$ V with a nominal 
operating temperature of 20$^{\circ}$C.  The energy scale calibration 
is performed automatically on the front end electronics (XA1) by injecting 
calibration pulses.  This electronic calibration task is executed every $\sim$ 
5000 s during spacecraft slews.  In addition to the electronic 
calibration, there are two $^{241}$Am tagged sources mounted below the mask 
for calibrating the absolute energy scale and the detector efficiency 
for each CZT pixel in-flight.  

There are three types of triggers in the BAT flight software.  Two of these 
are rate triggers looking for excesses in the count rate from the 
background, and one is an image trigger based on new significant sources 
found in the sky images.  
The rate triggers are divided into short 
(foreground period $\le$ 64 ms) and long rate triggers (foreground period 
$\ge$ 64 ms).  Each trigger criterion is a specific combination
of choices of foreground interval, number of background samples, energy
band and one of nine different regions of the detector plane.  Currently, 
494 trigger criteria have been running on-board for a long rate trigger, 36 for the 
short rate trigger, and 1 for the image trigger.  Once a rate trigger 
has occurred, the BAT creates a sky image using the triggered foreground 
and background intervals in the specified energy band to find a significant
source in the image.  Failing to produce a significant image excess, the BAT will
check for trigger criteria that produce a more statistically significant image.  
When a rate or image trigger finds a significant source in the image, 
a location match to the on-board source catalog is executed to 
exclude activity from known hard X-ray astronomical sources.\footnote
{If the significance of the activity of the known source is higher 
than the threshold value set in the on-board catalog, the activity 
will be reported to the GCN in real time with a different GCN notice type.}  

For each GRB trigger, the photon-by-photon data (event data) are
available with a time resolution of 100 $\mu$s.  The duration of 
the event data was initially from T-300 s to T+300 s (T as the BAT 
trigger time).  After March 17 2006, we have extended the duration 
of the event data which are downlinked to from T-300 s to T+1000 s.  
We also transmit 10 s 
of the event data for failed triggers.\footnote{Failed triggers are those 
which caused a rate trigger, but did not also have a significant point source 
in the image formed with that data; or those for which the image coincided with 
a known source in the on-board catalog.}  In the survey mode, 
the BAT produces detector plane histograms (DPHs).  These histograms 
have an 80 channel spectrum for each detector integrated typically over 
five minutes.\footnote{The integration time of DPHs changes in various
operational conditions, but is 300 s for most of the 
on-orbit operation time.}  The DPHs are the primary data product when 
the GRB prompt emission lasts longer than the stop time of the event 
data collection \citep[e.g. GRB 060124;][]{romano2006}.   

\section{On-orbit calibration}

The Crab nebula data collected for various positions in the BAT field of 
view were used for the on-orbit calibration.  We analyzed the 
DPH data for this purpose.  The standard BAT software (HEASoft 6.2) and the latest 
calibration database (CALDB: 2006-10-14) were used for processing the data.  
We first made the Good Time Interval (GTI) file 
for each observation segment (each observation ID) excluding the periods 
when 1) the spacecraft was not settled, 2) the spacecraft was in the South Atlantic 
Anomaly (SAA), and 3) the Crab was occulted by the earth, 
moon and/or sun.  
{\tt batoccultgti} was used for excluding the time periods for 
case 3.  
{\tt baterebin} was applied to the DPH data to correct the energy scale.  
A Detector Plane Image (DPI) 
file was created from the DPH using {\tt batbinevt} for each individual 
row.  The spacecraft attitude file was re-created using ftools {\tt aspect} by 
specifying the observation start and stop time of the DPI.  The detector 
enable/disable map was created using {\tt bathotpix}.  The BAT sky image 
was created by {\tt batfftimage} using the DPI, the updated attitude file and 
the enable/disable map.  {\tt batcelldetect} was used to extract the position of 
the Crab.  The mask weight map of the Crab was created by {\tt batmaskwtimg} 
using the ``true'' Crab position from SIMBAD (R.A.$_{\rm J2000}$ = 83.6332, 
Dec.$_{\rm J2000}$ = 22.0144).  
The spectrum (PHA file) was created by {\tt batbinevt} using the mask weight map for each 
row of the DPH.  All of the individual PHA files at the same sky coordinate 
were added to create a single PHA file if the Crab was detected $>8 \sigma$ in 
the image in the full energy band (15--350 keV).  The detector energy response 
file was created by {\tt batdrmgen} for each summed PHA file.  

\subsection{Position Accuracy}

\begin{figure}[p]
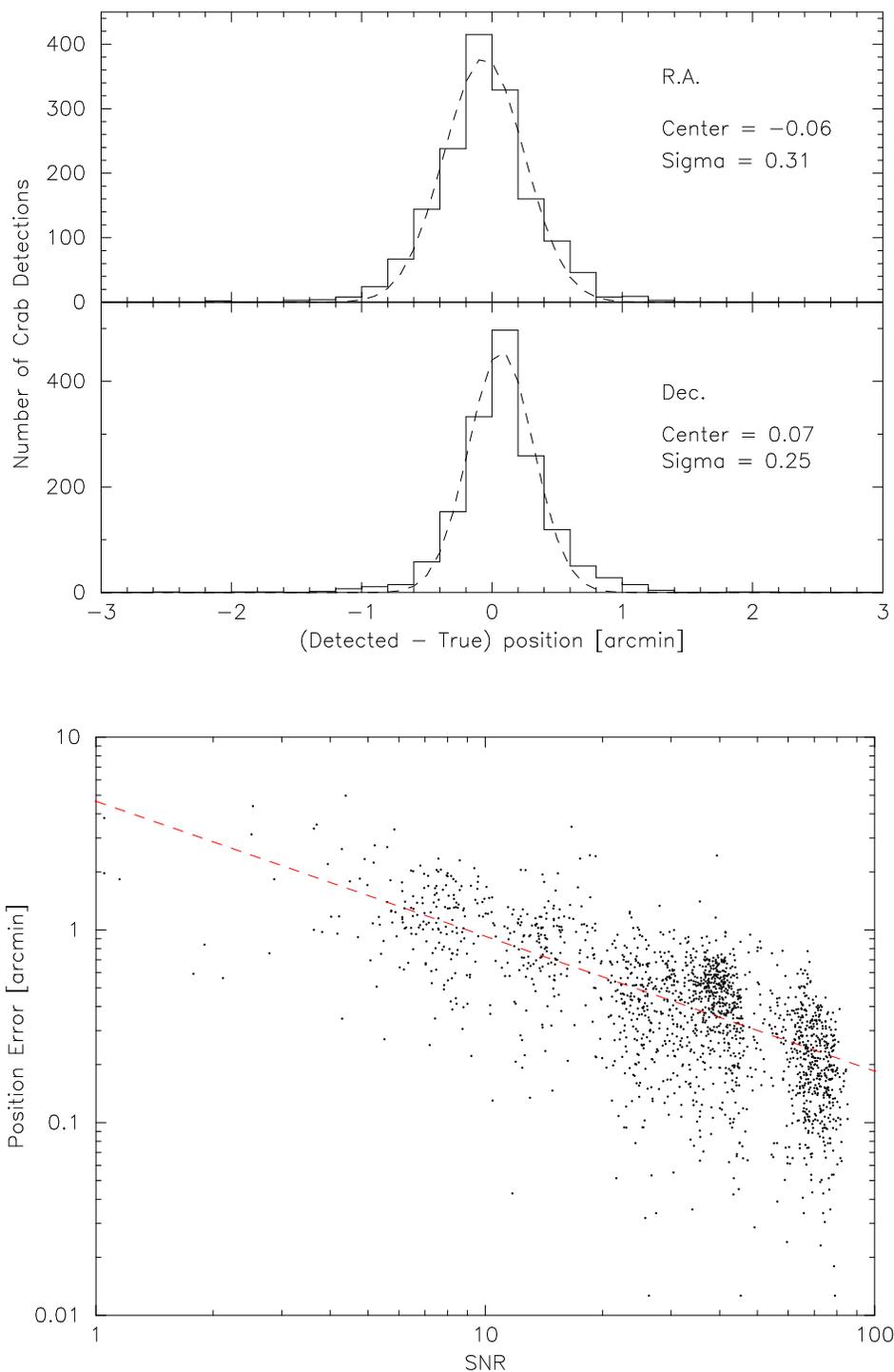

\centerline{
\includegraphics[width=9cm,angle=-90]{f1a.eps}}
\vspace{1cm}
\centerline{
\includegraphics[width=9cm,angle=-90]{f1b.eps}}
\caption{Top: The difference in R.A. (top) and Dec. (bottom) between
 the BAT detected ($> 8 \sigma$) and the SIMBAD ('true') Crab position.  The dotted
 line is the best fit gaussian model.  The centroid and sigma of the 
best fit gaussian are $-0.07^{\prime}$ and $0.33^{\prime}$ for 
R.A., and $0.08^{\prime}$ and $0.27^{\prime}$ for Dec., respectively.  Bottom:
BAT position errors as a function of signal to noise ratio for the Crab.  The red dashed 
line is the position error $\propto$ SNR$^{-0.7}$.}
\label{fig:crab_pos_accuracy}
\end{figure}

The histograms of the angular differences between the Crab detected position by 
{\tt batcelldetect} and the ``true'' Crab position (the Crab position in SIMBAD) 
are shown in the top of Figure \ref{fig:crab_pos_accuracy}.  The 
position differences are less than 1$^{\prime}$ for 95\% of the Crab 
observations in both the right ascension (R.A.) and the declination (Dec.).  
The bottom of Figure \ref{fig:crab_pos_accuracy} shows the BAT position errors as a 
function of signal to noise ratio for the Crab.  The signal to noise ratio 
and the position error are correlated with a power-law index of $-0.7$ (see section 5).  

\subsection{Energy Response}

\begin{figure}[p]
\centerline{
\includegraphics[width=10cm,angle=-90]{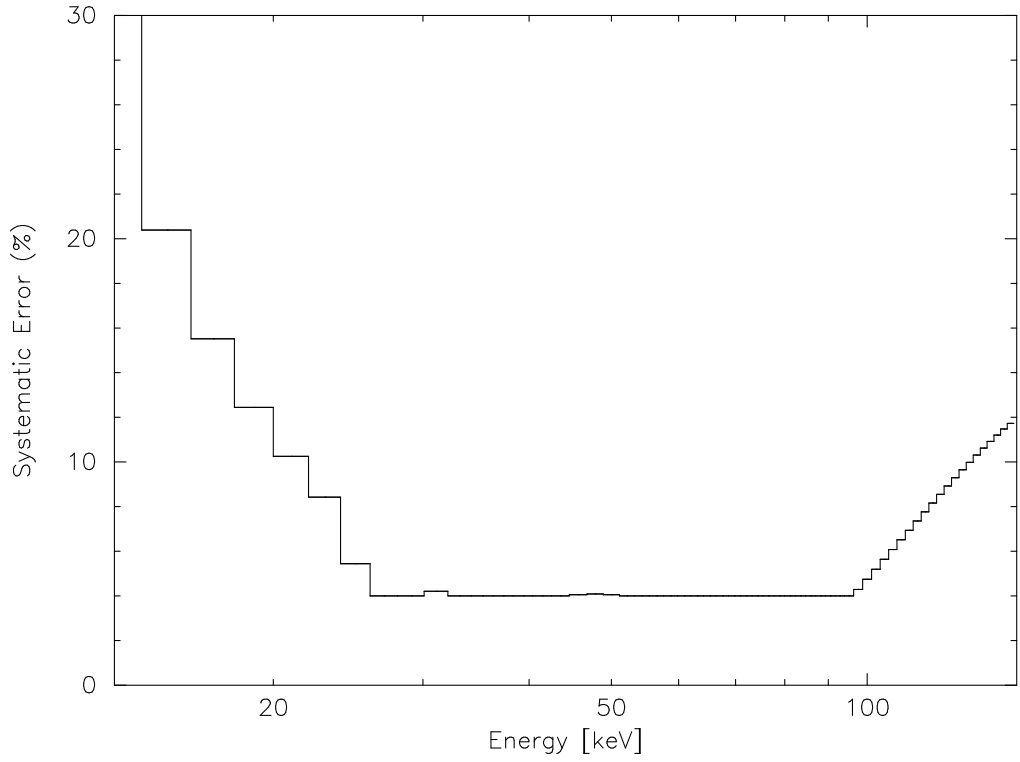}}
\caption{Systematic error as a function of energy.  The systematic
error vectors must be applied to the BAT spectral file created by 
the BAT software, HEASoft 6.2 and CALDB: 2006-10-14, due to the 
current uncertainly in the energy calibration.}
\label{fig:sysvector}
\end{figure}

Immediately after the first attempt to fit the Crab spectrum with the pre-launch 
detector energy response matrices (DRM), we noticed that there were systematic 
errors in the pre-launch DRM at low energies (below 25 keV) and also at 
high energies (above 80 keV).  The investigation of these problems is still 
in progress.  To overcome these problems, we applied corrections to force 
the Crab to fit a canonical model, a power-law with a photon index of 2.15 and 
a 15-150 keV energy flux of $2.11 \times 10^{-8}$ ergs cm$^{-2}$ s$^{-1}$ 
\citep[e.g.,][]{jung1989,rothschild1998,sax1999,hete2003,integral2005}.
Due to these corrections, 
the BAT team has released the software tool, {\tt batphasyserr}, and the 
CALDB file (swbsyserr20030101v002.fits) to apply the energy dependent 
systematic errors to the PHA file.  The systematic errors which should be 
applied to the PHA file are shown in Figure \ref{fig:sysvector}.  The Crab 
spectra were fitted by a simple power-law model using {\tt Xspec 11.3.2}
including the systematic errors.  

\begin{figure}[p]
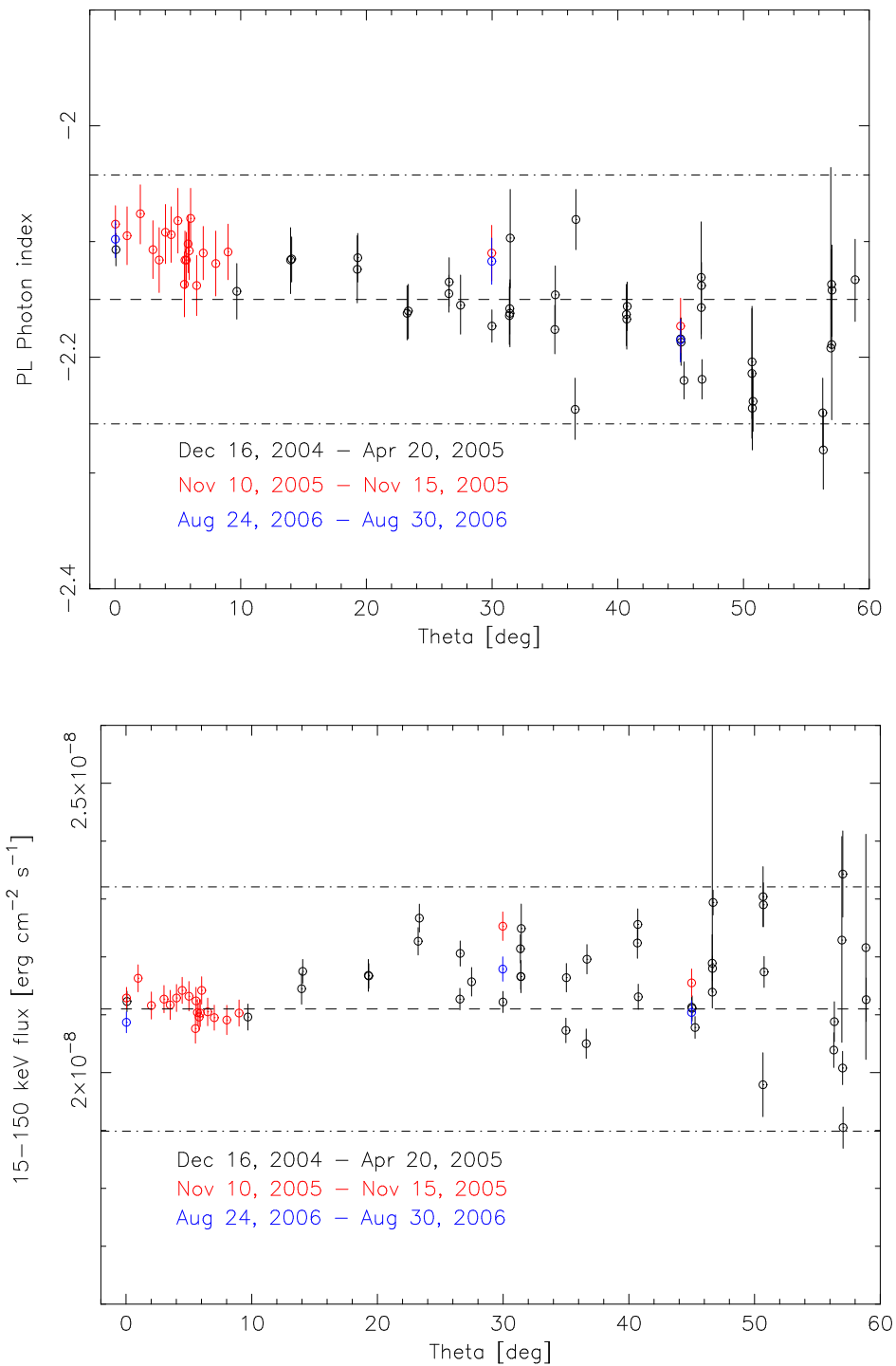

\centerline{
\includegraphics[width=9cm,angle=-90]{f3a.eps}}
\vspace{1cm}
\centerline{
\includegraphics[width=9cm,angle=-90]{f3b.eps}}
\caption{The power-law photon index (top) and the flux in the 
 15-150 keV band as a function of the incident angle ($\theta$) of the Crab 
observed December 2004-April 2005 (black), November 2005 (red), and
 August 2006 (blue).  The horizontal dashed lines are the Crab canonical
 values of $-2.15$ for the photon index and $2.11 \times 10^{-8}$ ergs
 cm$^{-2}$ s$^{-1}$ for the flux.  The dashed dotted lines are $\pm$5\% of 
the photon index and $\pm$10\% of the flux canonical values.}
\label{fig:crab_phindex_flux_theta}
\end{figure}

\begin{figure}[p]
\centerline{
\includegraphics[width=18cm,angle=0]{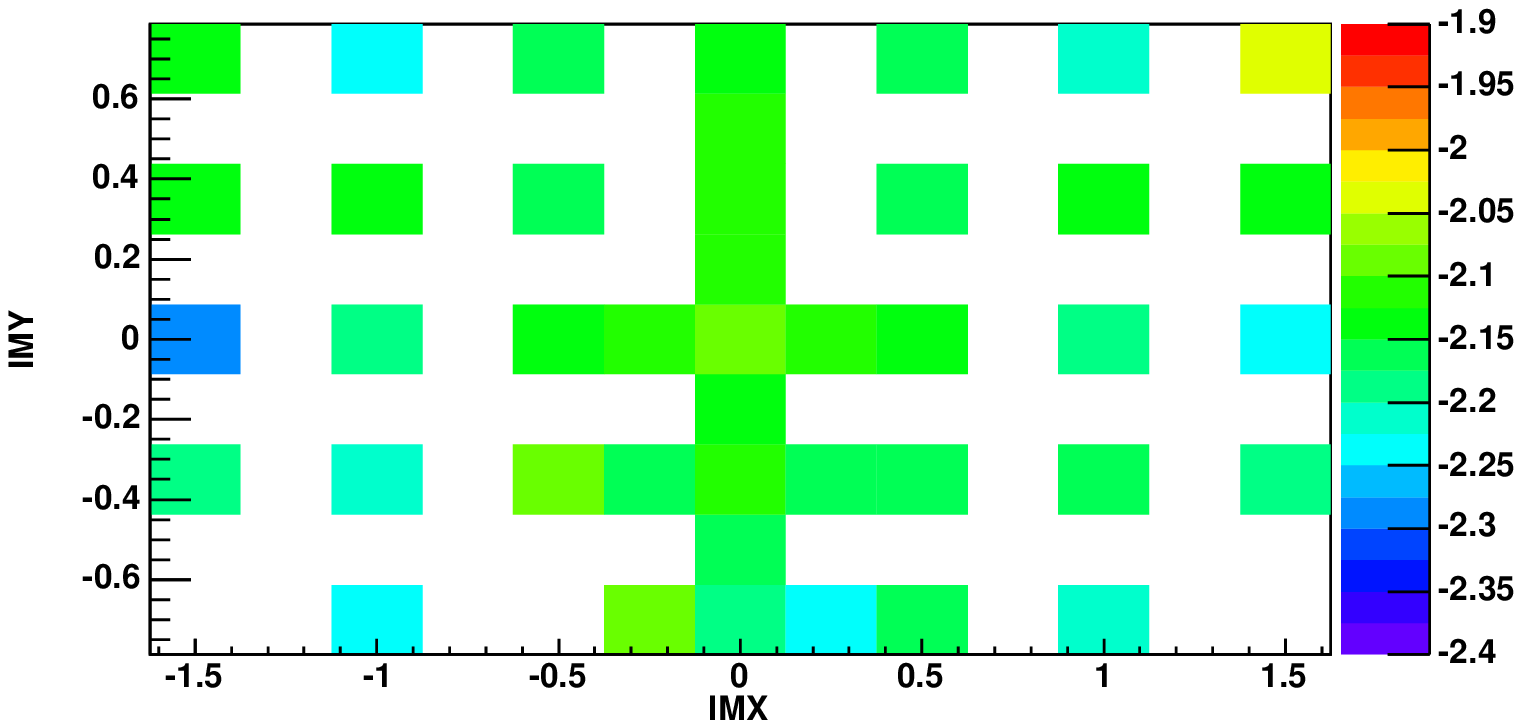}}
\vspace{1cm}
\centerline{
\includegraphics[width=18cm,angle=0]{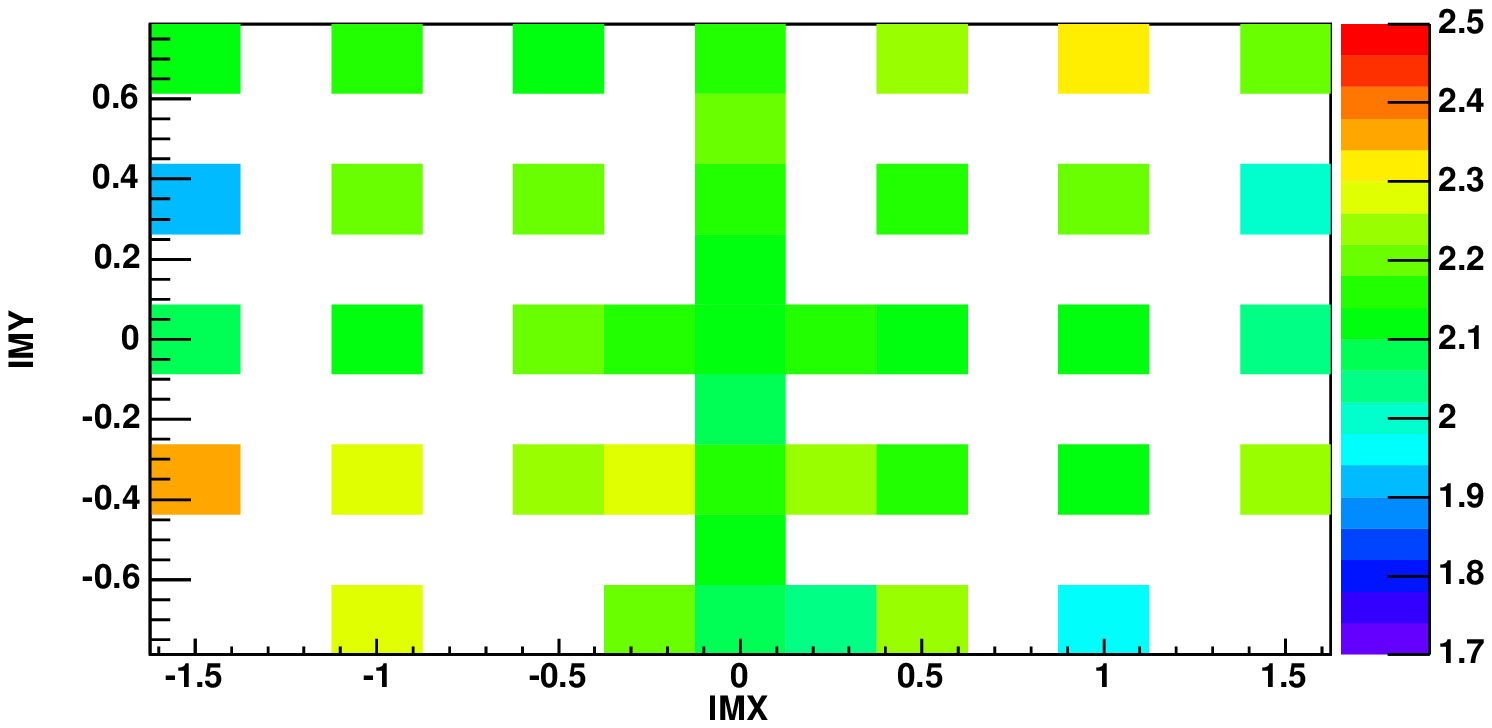}}
\caption{The contour maps (sparsely sampled) of the Crab photon index (top) and the flux in the
 15-150 keV band in units of 10$^{-8}$ ergs cm$^{-2}$ s$^{-1}$
 (bottom) in the BAT field of view in tangent plane coordinates (IMX and IMY).}
\label{fig:2d_crab_phindex_flux}
\end{figure}

Figure \ref{fig:crab_phindex_flux_theta} shows the Crab photon index and the 
flux in the 15-150 keV band as a function of the incident angle.  In the current 
BAT DRM ({\tt batdrmgen v3.3} and CALDB: 20061014), the scatter of the
photon index and the flux are about 5\% and 10\% of the canonical
values.  
Figure \ref{fig:2d_crab_phindex_flux} shows the contour maps of the Crab photon 
index and the flux in the 15-150 keV band over the BAT field of view.  
We note that the parameters tend to deviate 
from the canonical values towards 
the edge of the BAT field of view.  Thus, the spectral 
parameters could have a larger systematic error when the source is 
at the edge of the field of view of BAT.  

\subsection{BAT GRB Response Time}

\begin{figure}[p]
\centerline{
\includegraphics[width=10cm,angle=-90]{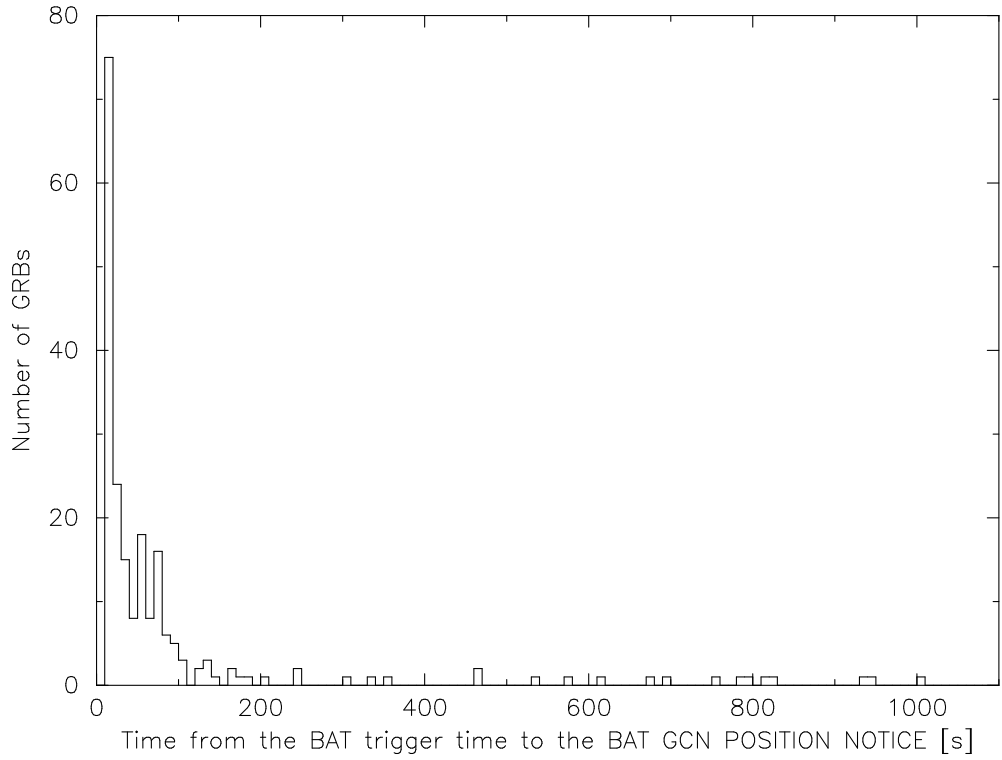}}
\caption{The time delay from the BAT trigger time to the GCN BAT Position Notice 
(the BAT burst sample from GRB 050215A to GRB 070616 excluding the GRBs found 
in ground processing).}
\label{fig:bat_pos_notice_time}
\end{figure}

Figure \ref{fig:bat_pos_notice_time} shows the histogram of the time delay 
between the BAT GRB trigger time and the GCN BAT Position Notice.\footnote{This GRB sample 
is from GRB 050215A to GRB 070616 (slightly reduced from the rest of the paper) 
excluding the GRBs found on the ground process.} 
The highest peak of the distribution is around 15 s.  The BAT position has been 
reported on the ground within 30 s after the burst trigger for half of the BAT GRBs.  
Most of the longer delays ($>$300 s) are due to interruptions in TDRSS transmissions 
during regular telemetry down links to the Malindi ground station.  

\section{Analysis for the GRB catalog}

We used the standard BAT software (HEASoft 6.1.1) and the latest
calibration database (CALDB: 20061014) to process the BAT GRBs 
from December 2004 (GRB 041217) to June 2007 (GRB 070616).\footnote{The GRB sample 
includes bursts which were found in the ground processing.}  
The burst pipeline script,
{\tt batgrbproduct}, was used to process the BAT event data.\footnote{By default, the minimum partial coding setting was 10\% to
 remove portions of the light curve with poor sampling, and the
 aperture setting was CALDB:FLUX in order to avoid passive materials
 in the BAT field of view.  Some bursts were initially in the extreme
 partial coded field of view ($<$10\%).  In those cases, we
re-ran {\tt batgrbproduct} specifying the options 
pcodethresh=0.0 and aperture=CALDB:DETECTION.}  
The {\tt Xspec} spectral fitting tool (version 11.3.1) was used to fit
each spectrum.  
Since our analysis is restricted to use only the event data, 
we present partial analysis based on the available 
event data for bursts which last longer than the end period of the event data 
(e.g. GRB 060124) or which have incomplete event data due to the various 
reasons (e.g. GRB 050507).  In some cases, especially for weak short GRBs, {\tt battblocks}, 
which is one 
of the task run in {\tt batgrbproduct}, might fail to find the burst interval.  
In those cases, 
we fitted the mask-weighted light curve in the full BAT energy range by a 
liner-rise exponential decay model (``BURS'' model in ftools 
qdp\footnote{http://heasarc.gsfc.nasa.gov/docs/software/ftools/others/qdp/qdp.html})
to find the burst time intervals (T$_{100}$, T$_{90}$, T$_{50}$ and peak 
1-s intervals) and created the T$_{100}$ and peak 1-s PHA files based on 
these time intervals.  We put comments on Table \ref{tbl:bat_summary} for 
the bursts which have a problem in either the data or the processing.  

For the time-averaged spectral analysis, we use the time interval from
the emission start time to the emission end time (T$_{100}$ interval).  
Since the BAT energy response generator, {\tt batdrmgen}, performs the
calculation for a fixed single incident angle of the source, it will be
a problem if the position of the source is moving during the time interval 
selected for the spectral analysis because of the spacecraft slew.  
In this situation, we created the DRMs for each five-second 
period during the time interval taking into account the position 
of the GRB in detector coordinates.  We then weighted these DRMs 
by the five-second count rates and created the averaged DRM using {\tt addrmf}.
Since the spacecraft slews about one degree per second in response to a GRB
trigger, we chose five second intervals to calculate
the DRM for every five degrees.  

\begin{figure}
\centerline{
\includegraphics[width=10cm,angle=-90]{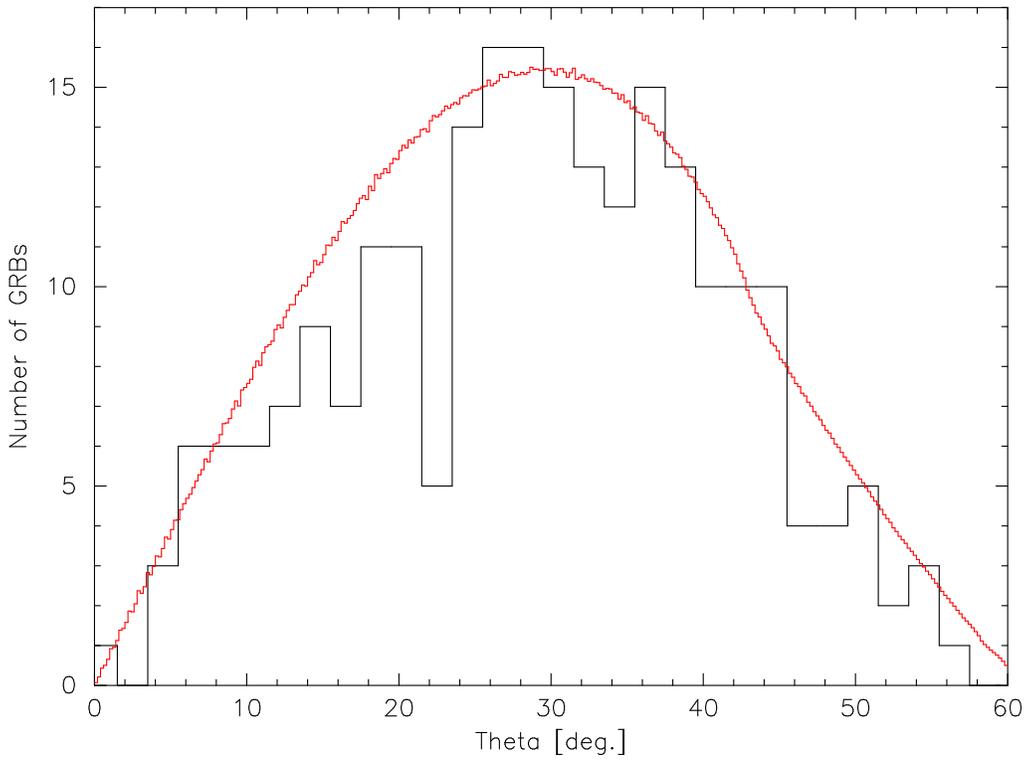}}
\caption{The incident angle ($\theta$) distribution of the BAT GRBs.  
The red line shows the calculated $\theta$ distribution in the case 
of the uniform sky distribution.}
\label{fig:bat_theta}
\end{figure}

The spectrum was fitted by a simple power-law (PL) model, 
\begin{eqnarray}
f(E) = K_{50}^{\rm PL}\left(\frac{E}{50  \: {\rm keV}}\right)^{\alpha^{\rm PL}} 
\label{eq:pl}
\end{eqnarray}
where $\alpha^{\rm PL}$ is the power-law photon index and $K_{50}^{\rm
PL}$ is the normalization at 50 keV in units of 
photons cm$^{-2}$ s$^{-1}$ keV$^{-1}$, and by a cutoff power-law (CPL) model, 
\begin{eqnarray}
f(E) = K_{50}^{\rm CPL}\left(\frac{E}{50 \: {\rm keV}}\right)^{\alpha^{\rm CPL}} 
\exp\left(\frac{-E\,(2+\alpha^{\rm CPL})}{\ep}\right)
\label{eq:cpl}
\end{eqnarray}
where $\alpha^{\rm CPL}$ is the power-law photon index, $\ep$ is the
peak energy in the $\nu$F$_{\nu}$ spectrum and $K_{50}^{\rm CPL}$ is the 
normalization at 50 keV 
in units of photons cm$^{-2}$ s$^{-1}$ keV$^{-1}$.  We also systematically 
fitted the spectrum with the Band function \citep{band1993}.  However, none of the 
BAT spectra show a significant improvement in $\chi^{2}$ with 
a Band function fit compared to that of a CPL model fit.  
Note that this is equivalent to saying that a CPL model and a Band function 
represent equally well the observed spectrum.  However, we only present 
the results based on 
a CPL model throughout the paper due to its simplicity in the functional 
form.\footnote{A Band function has 4 parameters, whereas, a CPL model 
has 3 parameters.}  
The best fit spectral model is determined 
based on the difference in $\chi^{2}$ between a PL and a CPL fit.  If
$\Delta$ $\chi^{2}$ between a PL and a CPL fit is greater than 6 ($\Delta \chi^{2}
\equiv \chi^{2}_{\rm PL} - \chi^{2}_{\rm CPL}$ $>$ 6), we determined that a
CPL model is a better
representative spectral model for the data.  To quantify the significance of this
improvement, we performed 10,000 spectral simulations taking into account the
distributions of the power-law photon index in a PL fit, the fluence in the 15-150 keV band
in a PL fit and the T$_{100}$ duration of the BAT GRBs, and determined
how many cases a CPL fit gives
$\chi^{2}$ improvements of equal or greater than 6 over a PL fit.  The BAT
DRM used in the simulation was for an incident angle of 30$^{\circ}$ 
which was an averaged incident angle of the BAT GRB sample 
(see Figure \ref{fig:bat_theta}).  We found equal or higher 
improvements in $\chi^{2}$ in 62 simulated spectra out of 10,000.  Thus,
the chance
probability of having an equal or higher $\Delta \chi^{2}$ of 6 with a CPL model when
the parent distribution is a case of a PL model is 0.62\%.

The fluence and the peak flux are derived from the spectral fits.  The
fluences are calculated fitting the time-averaged spectrum by the best 
fit spectral model.  The peak fluxes are calculated fitting the spectrum 
of the one-second interval bracketing the highest peak in the light 
curve (hereafter peak spectrum).  Again, we used the best fit spectral
model for calculating the peak fluxes.  To correctly reflect the incident 
angle of the source during the period of the peak spectrum, the DRM 
for the peak spectrum was created updating the keywords of the peak 
spectrum file by {\tt batupdatephakw} and running {\tt batdrmgen} using 
this updated spectral file.  

\section{The Catalog}

The first BAT catalog includes the time period between 2004 December 
19 and 2007 June 16.  The total number of GRBs including five untriggered 
GRBs\footnote{GRBs which found in ground processing.} and four 
possible GRBs is 237.  237 GRBs are listed in Table 
\ref{tbl:bat_summary}.  The first column is the GRB name.  The next column 
is the BAT trigger number.  The next column specifies the BAT trigger time 
in UTC in the form of {\it YYYY-MM-DD hh:mm:ss} where {\it YYYY} 
is year, {\it MM} is month, {\it DD} is day of month, {\it hh} is hour, 
{\it mm} is minute, and {\it ss} is second.  Note that the definition 
of the BAT trigger time is the start time of the foreground time interval 
of the image from which the GRB is detected on-board.  The next four columns give 
the locations by the ground process in equatorial (J2000) coordinate, 
the signal-to-noise ratio of the BAT image at the location, and the radius 
of the 90\% confidence region in arcmin.  The 90\% error radius is calculated 
based on the signal-to-noise ratio of the image using the following equation 
which derived from the BAT hard X-ray survey 
process\footnote{http://heasarc.gsfc.nasa.gov/docs/swift/analysis/bat\_digest.html}$^{,}$\footnote{SWIFT-BAT-CALDB-CENTROID-v1 \\
http://swift.gsfc.nasa.gov/docs/heasarc/caldb/swift/docs/bat/index.html}
\begin{displaymath}
r_{90\%} = 10.92 \times {\rm SNR}^{-0.7}\;({\rm arcmin}), 
\end{displaymath}
where SNR is the signal-to-noise ratio of the BAT image.  However, due to the 
limitation of the BAT point spread function, we decided to quote the minimum 
allowed value of $r_{90\%}$ as 1$^{\prime}$ in the catalog.  
The next two columns specify the 
burst durations which contain from 5\% to 95\% (T$_{90}$) and from 25\% to 
75\% (T$_{50}$) of the total burst fluence.  These durations are calculated in 
the 15--350 keV band.\footnote{The coded mask is transparent to photons above 150 keV.
Thus, photons above 150 keV are treated as background in the mask-weighted
method.  The effective upper boundary is $\sim$ 150 keV.}  The next two columns are 
the start and stop time from the BAT trigger time of the event data.  The last column is 
the comments.  

The energy fluences calculated in various energy bands are summarized in 
Table \ref{tbl:bat_fluence}.  The first column is the GRB name.  The next column specifies 
the spectral model which used in deriving the fluences (PL: simple power-law model; 
Eq.(\ref{eq:pl}), CPL: cutoff power-law model; Eq.(\ref{eq:cpl})).  The next five columns 
are the fluences in the 15-25 keV, the 25-50 keV, the 50-100 keV, the 100-150 keV, and the 
15-150 keV band.  The unit of the fluence is 10$^{-8}$ ergs cm$^{-2}$.  
The last two columns specify the start and the stop time from the 
BAT trigger time which used to calculate the fluences.  Note that since our analysis is 
based on the available event data, 6 bursts with the incomplete data (see the 12th column 
of Table \ref{tbl:bat_summary}) might not include the whole burst emission.  

Table \ref{tbl:bat_phflux} and \ref{tbl:bat_peakeneflux} 
summarize the 1-s peak photon and energy fluxes in various energy 
bands.  The first column is the GRB name.  The next column specifies 
the spectral model used in deriving the 1-s peak flux.  The next five column 
are the 1-s peak photon and energy fluxes in the 15-25 keV, the 25-50 keV, the 50-100 keV, 
the 100-150 keV, and the 15-150 keV band.  The unit of the flux is 
photons cm$^{-2}$ s$^{-1}$ for the peak photon flux and 10$^{-8}$ ergs cm$^{-2}$ s$^{-1}$ 
for the peak energy flux.  
The last two columns specify the start and the stop time from the BAT trigger time 
which were used to calculate the peak fluxes.  

The time-averaged spectral parameters are listed in Table \ref{tbl:bat_spec}.  
The first column is the GRB name.  The next three columns are the photon index, 
the normalization at 50 keV and $\chi^{2}$ of the fit for a PL model.  The degree of 
freedom is 57 for all bursts in a PL fit.  The next four columns are the photon 
index, $\eop$, the normalization at 50 keV and $\chi^{2}$ of the fit in a CPL model.  
The degree of freedom is 56 for all bursts for a CPL fit.  The spectral parameters in 
a CPL are only shown for the bursts which meet the criteria described in the section 4.  

In the following subsections of investigating the relationship among fluences, 
peak fluxes, and the spectral parameters, we excluded 10 GRBs based on an incomplete 
data set and also those labeled as possible GRBs/SGRs  
(see the 12th column of Table \ref{tbl:bat_summary}).  

\subsection{BAT GRB Position and Sky Locations}

\begin{figure}
\centerline{
\includegraphics[width=10cm,angle=-90]{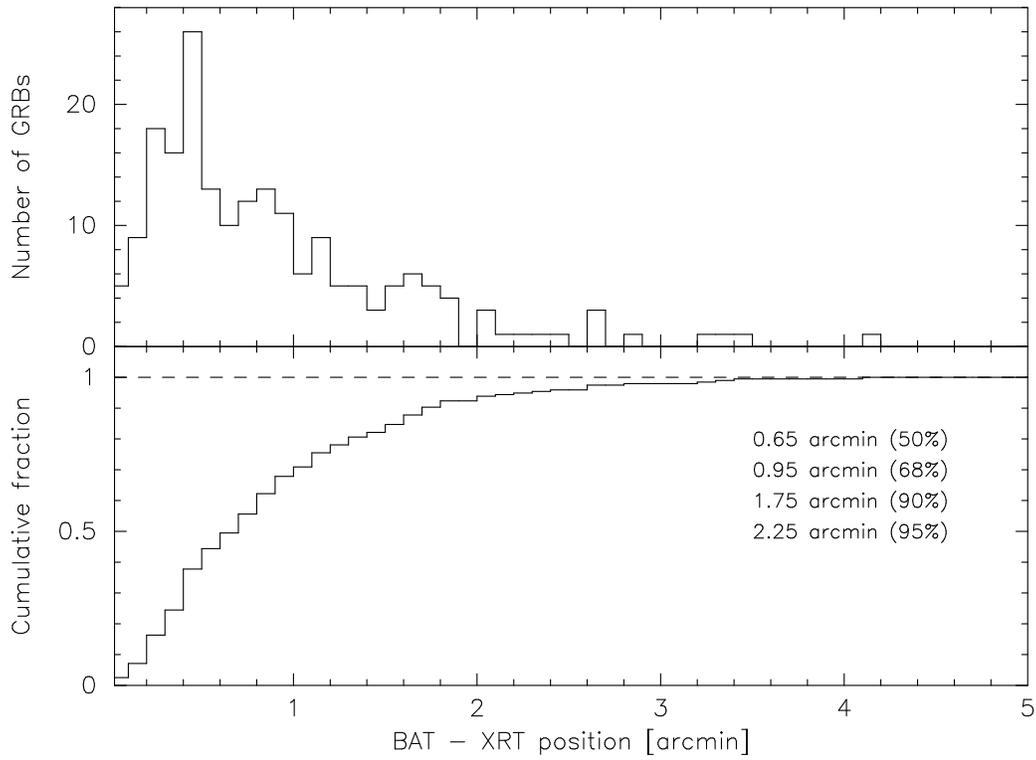}}
\caption{The histogram (top) and the cumulative fraction (bottom) 
of the angular difference between the BAT ground position and the 
XRT position.  68\% and 90\% of BAT ground positions are within 
0.95$^{\prime}$ and 1.75$^{\prime}$ from the XRT position, respectively.}
\label{fig:bat_xrt_pos_diff}
\end{figure}

\begin{figure}
\centerline{
\includegraphics[width=15cm,angle=0]{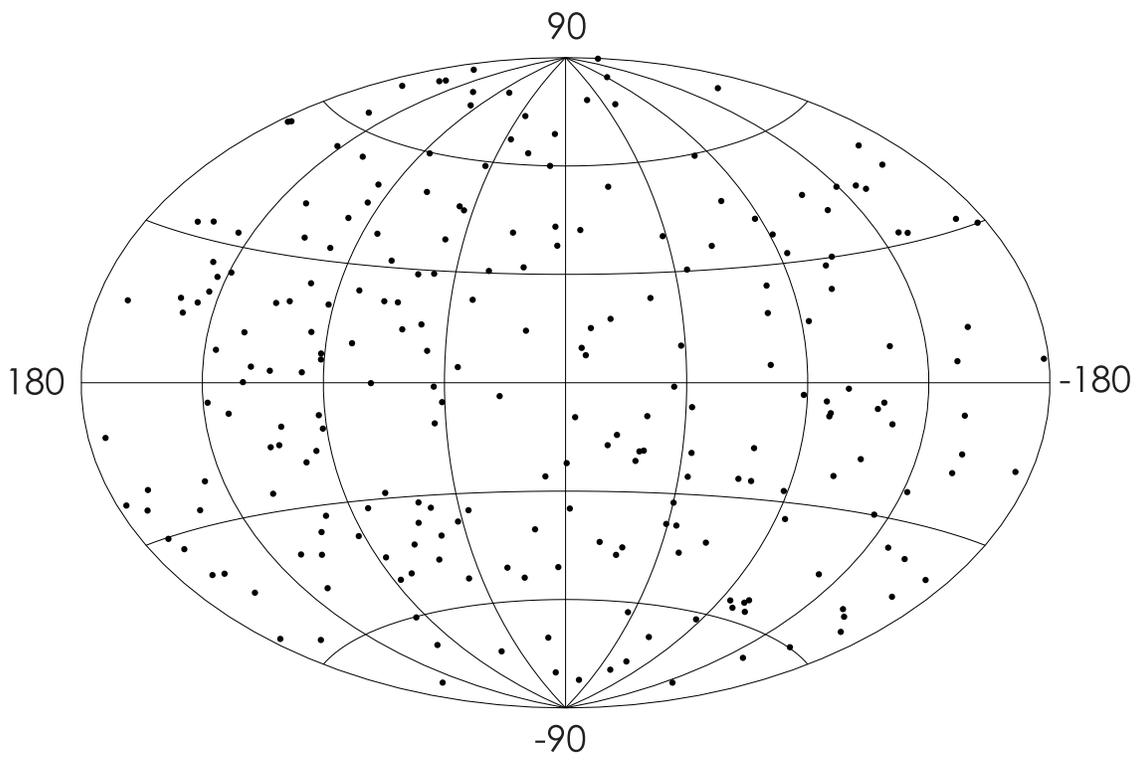}}
\caption{Sky distribution of the 237 BAT bursts in Galactic coordinates.}
\label{fig:bat_sky_map}
\end{figure}

Figure \ref{fig:bat_xrt_pos_diff} shows the angular difference between the 
BAT ground position and the first reported XRT position.\footnote{The XRT position 
is either from the on-board or from the XRT event-by-event data downloaded from the 
TDRSS satellite (Single-Pixel-Event-Report).}  The BAT ground position is within 
0.95$^{\prime}$ and 1.75$^{\prime}$ from the XRT position for 68\% and 90\% of the 
bursts, respectively.  The distribution of the incident angles ($\theta$) is shown in Figure 
\ref{fig:bat_theta}.  The $\theta$ distribution is peaked around 30$^{\circ}$ with a spread 
from 0$^{\circ}$ to 60$^{\circ}$.  The sky map of the BAT 237 GRBs in galactic 
coordinates is shown in Figure \ref{fig:bat_sky_map}.  

\subsection{Durations and Hardness}

\begin{figure}
\centerline{
\includegraphics[width=10cm,angle=-90]{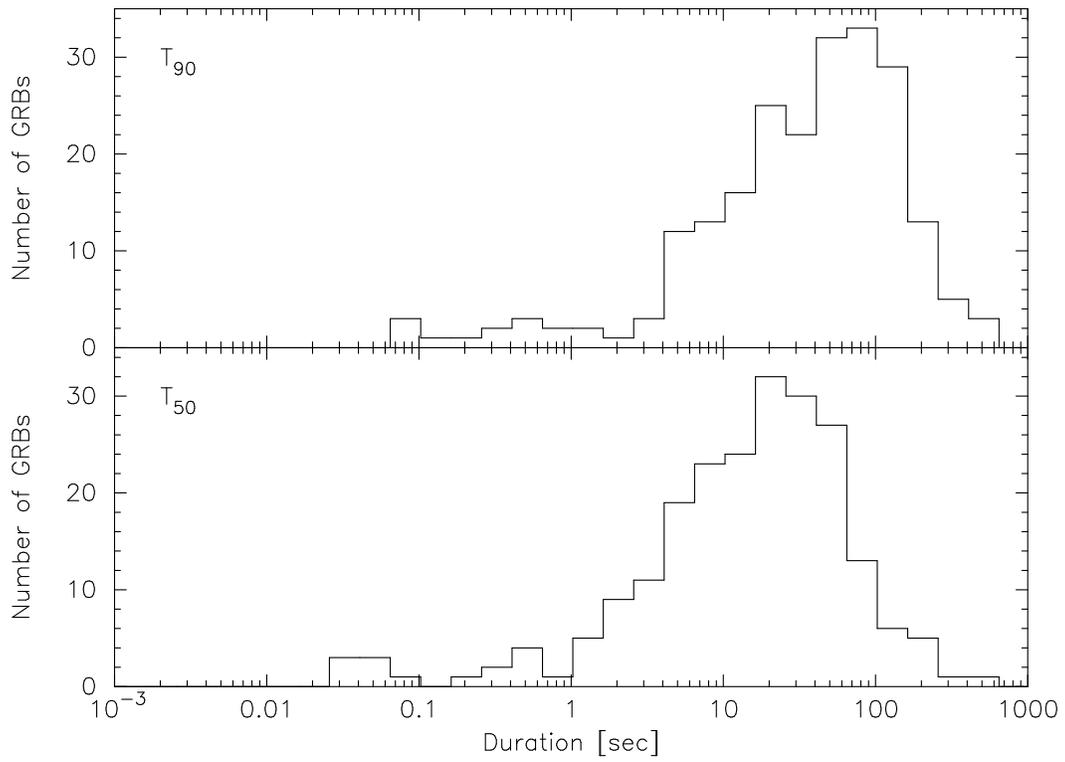}}
\caption{T$_{90}$ (top) and T$_{50}$ (bottom) distributions from the BAT
 mask-weighted light curves in the 15-350 keV band.}
\label{fig:bat_t90_t50}
\end{figure}

The histograms of T$_{90}$ and T$_{50}$ measured by the mask weighted 
light curve in the BAT full energy band 
are shown in Figure \ref{fig:bat_t90_t50}.  The BAT T$_{90}$ and T$_{50}$ 
durations are peaked around 80 s and 20 s respectively.  The BAT duration distributions 
do not show the clear bimodality seen in the BATSE sample \citep[e.g.,][]{kouveliotou1993} 
due to the smaller samples of short duration bursts (hereafter short GRBs).  
This is a well-known 
selection effect for an imaging GRB instrument like the BAT.  For most of the GRB imaging 
instruments, two triggering processes have to be met to be determined a GRB trigger.  
The first step is the increase in the count rate from the background level, and 
the second step is the significant signal in the image.  Although 
the short duration bursts would have triggered as an excess in the count rate, 
it could be very 
difficult to meet the criterion for a significant signal in the image 
because of the smaller number of photons 
to do the imaging compared to those of the long duration bursts (hereafter long GRBs).  
However, because of the 
large effective area and also the sophisticated flight software, the BAT has 
been triggering and localizing short GRBs in a much higher 
fraction than other GRB imaging instruments (e.g. {\it Beppo}SAX, {\it HETE-2}).  

\begin{figure}
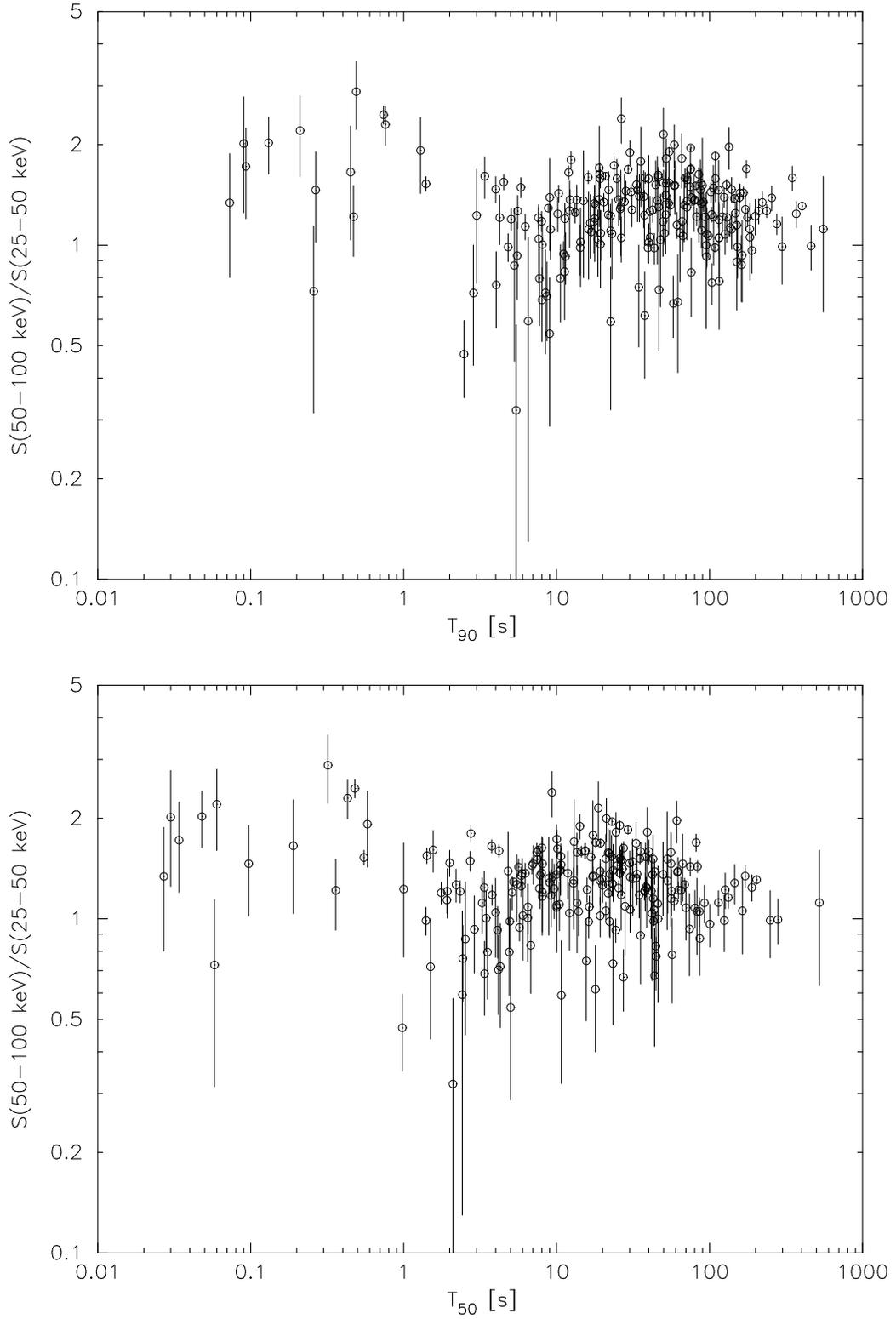

\centerline{
\includegraphics[width=10cm,angle=-90]{f10a.eps}}
\vspace{0.5cm}
\centerline{
\includegraphics[width=10cm,angle=-90]{f10b.eps}}
\caption{T$_{90}$ (top) and T$_{50}$ (bottom) versus the fluence ratio between
 the 50-100 keV and the 25-50 keV bands.}
\label{fig:bat_t90_t50_vs_s23}
\end{figure}

Figure \ref{fig:bat_t90_t50_vs_s23} shows the T$_{50}$ and T$_{90}$ durations 
versus the fluence ratio between the 50-100 keV and the 25-50 keV band.  Although 
the short GRBs tend to be systematically harder than the long GRBs, 
the separation in the hardness between these two classes is not 
obvious, at least in the BAT GRB sample.  Note that there are several 
works that question the hardness of the BATSE short GRBs 
reported on the BATSE catalog \citep{sakamoto2006,ohno2007,donaghy2007}.  
Therefore, as mentioned in \citet{donaghy2007} and also seen in the BAT 
short GRB samples, 
the duration and the 
hardness are insufficient to resolve the long and short populations, 
since the two components have so much overlap.  
\subsection{Peak Fluxes and Fluences}

\begin{figure}
\centerline{
\includegraphics[width=10cm,angle=-90]{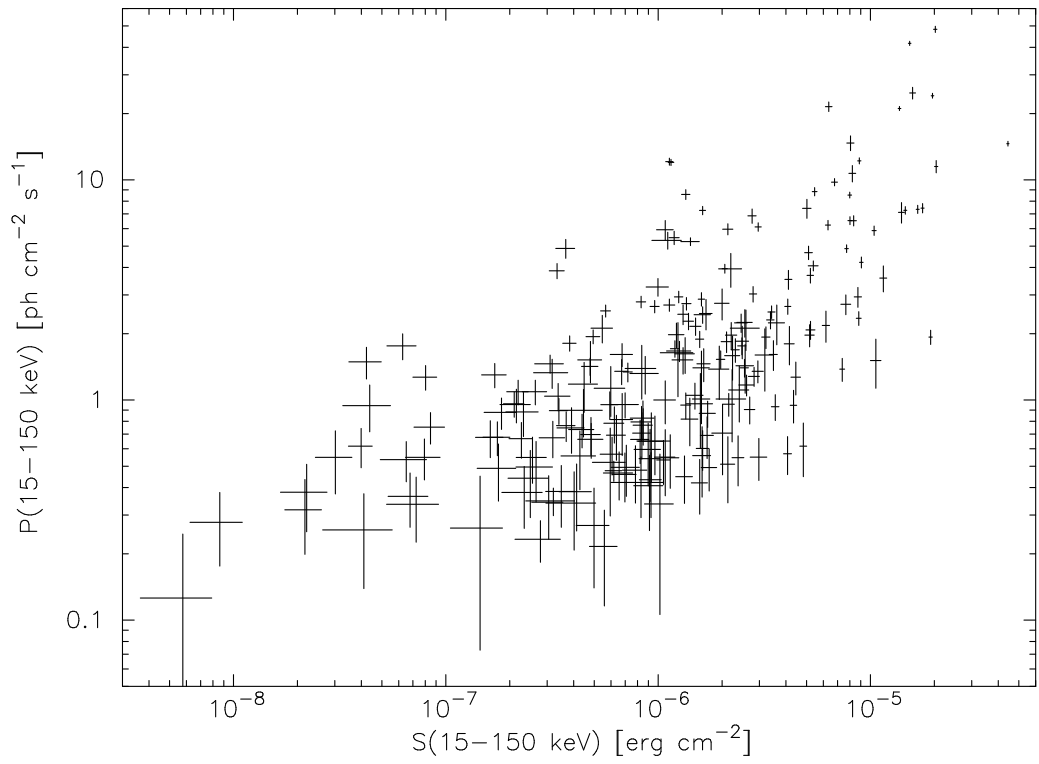}}
\caption{The distribution of the energy fluence in the 15-150 keV band
 versus 1-s peak photon flux in the 15-150 keV band.}
\label{fig:bat_peakflux_fluence}
\end{figure}

Figure \ref{fig:bat_peakflux_fluence} shows the distribution of 
the fluence versus the peak photon flux in the 15-150 keV band.  
The positive correlation is seen in these two parameters (correlation 
coefficient of +0.912 in a 220 burst sample).  
The peak photon flux of about 70\% of the BAT GRBs 
is less than 2 photons cm$^{-2}$ s$^{-1}$ in the 15-150 keV band.  If we 
assume a GRB detector with a sensitivity of 1 photons 
cm$^{-2}$ s$^{-1}$ in the 50-300 keV band (trigger band of BATSE), 
the peak photon flux in the 15-150 keV band is 2.5 photons 
cm$^{-2}$ s$^{-1}$ assuming the Band function parameters of $\alpha = -1$, 
$\beta = -2.5$, and $\eop = 100$ keV.  Thus, the majority of the BAT GRBs 
might be too weak to trigger a typical GRB detector which is sensitive at 
$>$50 keV.  This rough estimate 
is consistent with the observation that only 20-30\% of the BAT GRBs are 
simultaneously triggered successfully by the currently active GRB detectors 
such as ${\it Konus}$-${\it Wind}$ and ${\it Suzaku/WAM}$.  

\begin{figure}
\centerline{
\includegraphics[width=10cm,angle=-90]{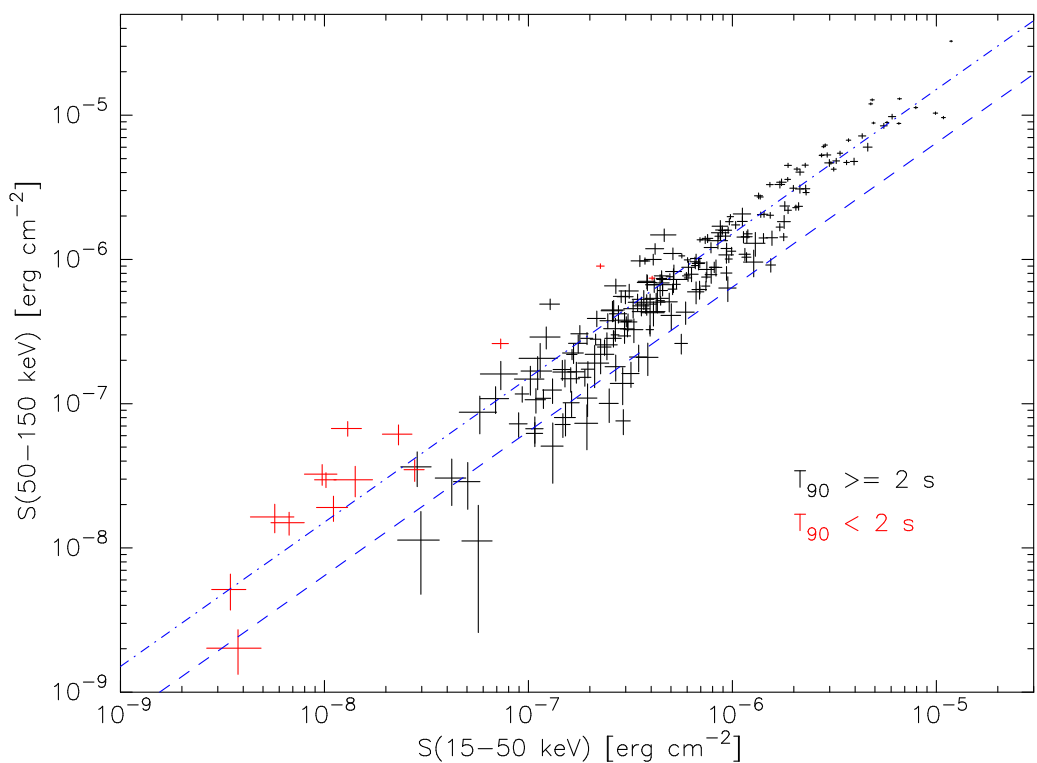}}
\caption{The distribution of the energy fluence in the 15-50 keV band
 versus that in the 50-150 keV band.  Long GRBs (T$_{90}$ $\ge$ 2 s) are in 
 black and short GRBs (T$_{90}$ $<$ 2 s) are in red.  The blue
 dash-dotted line is the case of the Band function of $\alpha=-1$, 
$\beta=-2.5$, and $\eop=100$ keV.  The blue dashed line is the case
 of the Band function of $\alpha=-1$, $\beta=-2.5$, and $\eop=30$ keV.}
\label{fig:s12_s34}
\end{figure}

\begin{figure}
\centerline{
\includegraphics[width=10cm,angle=-90]{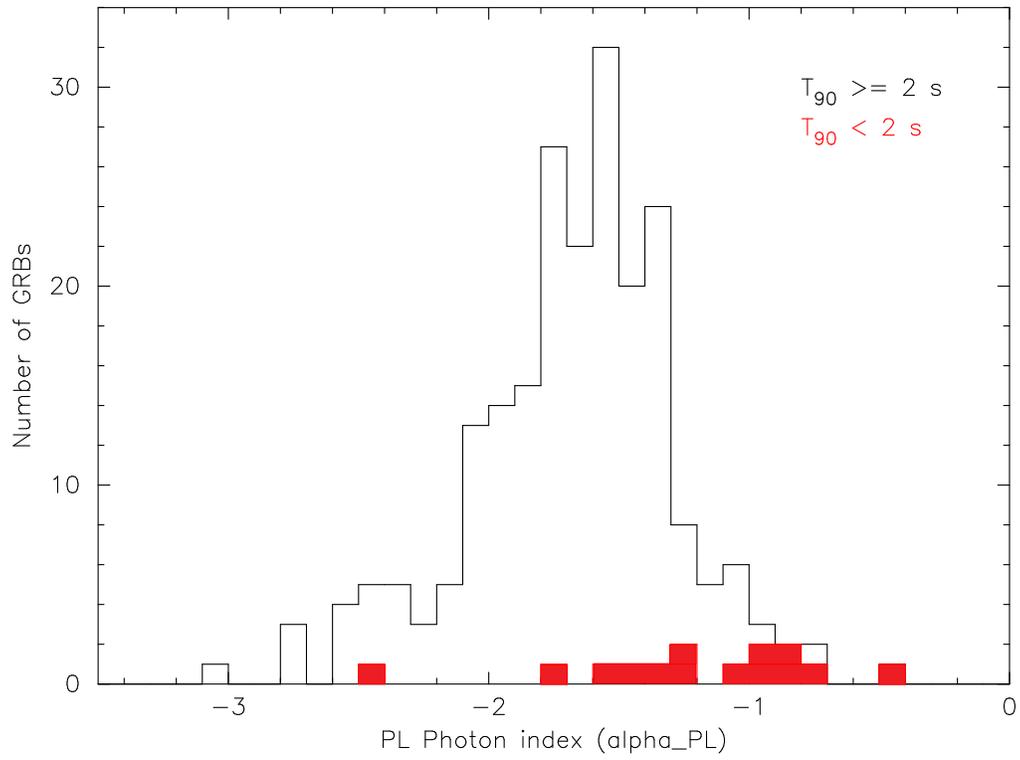}}
\caption{The histogram of the photon index in a PL fit for long GRBs
 (black) and short GRBs (red).  The short GRB which has a PL photon index 
of $-2.5$ is GRB 050906.}
\label{fig:pl_phoindex_hist}
\end{figure}

Figure \ref{fig:s12_s34} shows the fluence in the 15-50 keV band versus that in 
the 50-150 keV band.  The blue dash-dotted line is the case of the Band 
function parameters of $\alpha = -1$, $\beta = -2.5$, and $\eop = 100$ keV.  
The blue dotted line is the case of the Band 
function parameters of $\alpha = -1$, $\beta = -2.5$, and $\eop = 30$ keV.  
With the assumption of the Band parameters of $\alpha = -1$ and $\beta = -2.5$, 
the fraction of the long GRBs (T$_{90}$ $>$ 2 s) with $\eop < 100$ keV can 
be estimated to be about 60\% of the total long GRB samples.  
On the other hand, 
according to the BATSE spectral catalog \citep{kaneko2006}, 
there are only 3\% of the long BATSE GRBs with $\eop < 100$ keV. 
Therefore, the $\eop$ distribution of the BAT GRBs could be systematically 
lower than the BATSE $\eop$ distribution because of the relatively 
lower energy coverage of the BAT
\footnote{Note that the estimated number of {\it Swift} GRBs with 
$\eop < 100$ keV depends on the assumption of $\alpha$.  If we estimate the 
number of {\it Swift} GRBs of $\eop < 100$ keV for $\alpha = -1.6$, which 
is the peak of 
the PL photon index distribution of BAT (Figure \ref{fig:pl_phoindex_hist}), 
and $\beta =-2.5$, 
it will be 10\% of the total population.  However, based on the Band $\alpha$ 
distribution of the BATSE spectral catalog \citep{kaneko2006} and also on the 
detailed spectral 
simulation study of the BAT data \citep{sakamoto2007}, the BAT PL photon 
index very likely does not correspond to $\alpha$ of the Band function 
(see section 5.4).}.  
\subsection{Time-averaged Spectral Parameters}

The histogram of the photon index in a PL fit is shown in Figure 
\ref{fig:pl_phoindex_hist}.  The photon index distribution has a 
single broad peak which centroid around $-1.6$.  The peak value 
of the PL photon index of $-1.6$ is close to the mean value of the 
low energy photon index of $-1.0$ and the high energy 
photon index of $-2.5$ of the typical GRB spectrum 
\citep{kaneko2006,sakamoto2005}.  Since we would expect a photon 
index based on a PL model to be $-1.0$ or $-2.5$ if $\eop$ is above or below 
the BAT energy band, this result clearly demonstrates that majority 
of $\eop$ of the BAT GRBs is likely to be within the BAT energy band.  
The reason why the BAT can not measure $\eop$ for the majority of GRBs 
is due to its narrow energy band \citep{sakamoto2007}.  
Although the sample of bursts is very limited, the short GRBs 
tend to have a harder PL photon index than the long GRBs.  
However, the PL photon index distributions have a significant 
overlap between the short and long GRBs.  

\begin{figure}
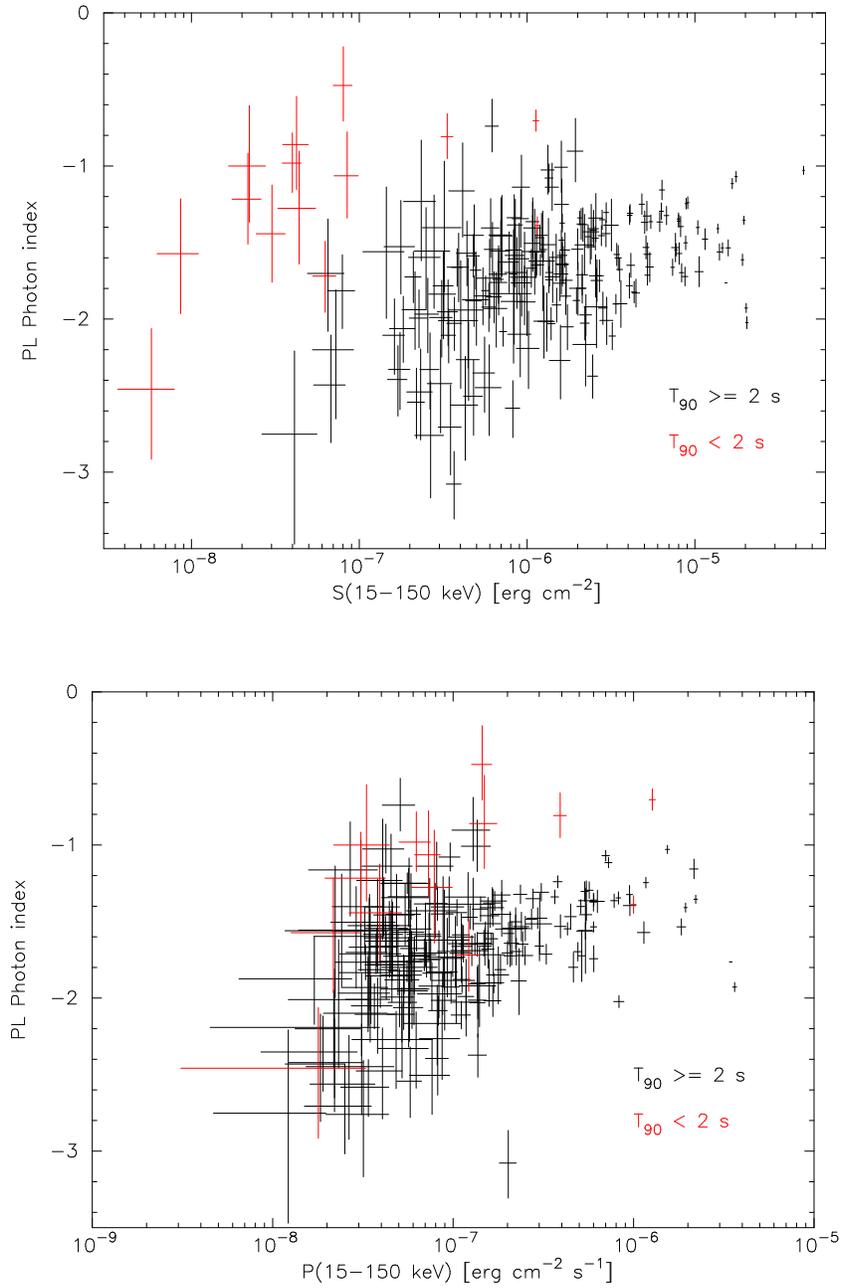

\centerline{
\includegraphics[width=8cm,angle=-90]{f14a.eps}}
\vspace{1cm}
\centerline{
\includegraphics[width=8cm,angle=-90]{f14b.eps}}
\caption{Top: The distribution of the PL photon index versus 
the energy fluence in the 15-150 keV band 
for long GRBs (black) and short GRBs (red). Bottom: The 
distribution of the PL photon index versus 
the 1-s peak energy flux in the 15-150 keV band 
for long GRBs (black) and short GRBs (red).}
\label{fig:fluence_pflux_phindex}
\end{figure}

The top panel of Figure \ref{fig:fluence_pflux_phindex} shows the photon 
index versus the fluence in the 15-150 keV band in a PL fit.  
There is a correlation between 
these two parameters for the long GRBs.  The correlation coefficient 
is +0.458 for the sample of 206 bursts.  The probability of such a correlation 
occurring by chance is $<$ 0.001\%.  If we include the short duration bursts, 
however, the correlation becomes weaker (correlation coefficient of +0.228 for the 
sample of 220 bursts).  Therefore, we have confirmed the fluence - hardness 
correlation \citep[e.g.,][]{lloyd2000} especially for the 
BAT long GRB sample. 
The bottom panel of Figure \ref{fig:fluence_pflux_phindex} shows the time-averaged 
photon index in a PL fit 
versus the 1-s peak energy flux in the 15-150 keV band.  The 
correlation coefficient of these two parameters are +0.397 for 
the GRB sample without the short bursts (total 206 GRBs) and +0.376 
for the sample with the short bursts (total 220 GRBs).  
The probabilities of a chance coincidence 
of the correlation between parameters are $<$ 0.001\% for both cases.  Therefore, 
we also confirmed the correlation between the peak flux and the hardness either 
with or without the short GRBs \citep[e.g.,][]{lloyd2000}.  Note that both 
the fluence and the peak flux measured by the BAT are not the bolometric values 
and may introduce the systematic errors in the correlations.    

\begin{figure}
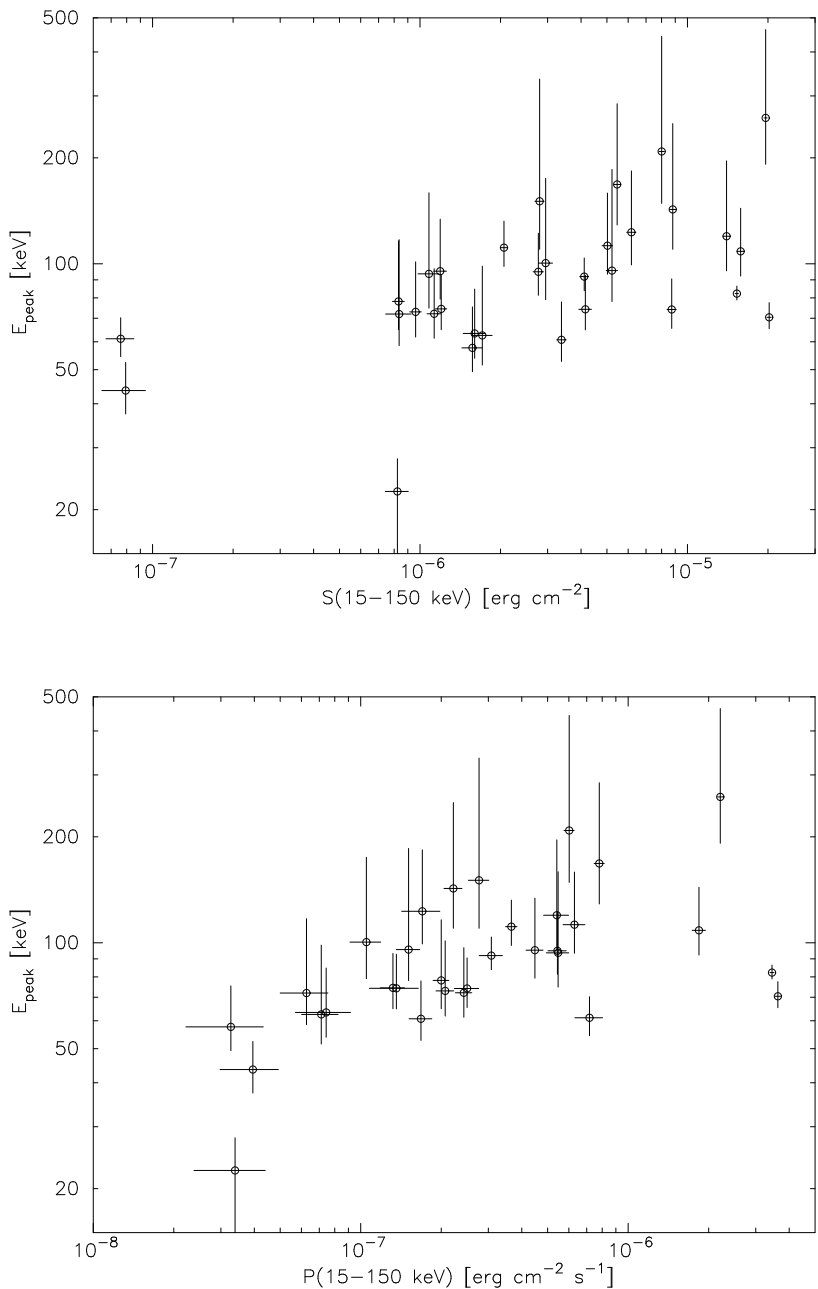

\centerline{
\includegraphics[width=8cm,angle=-90]{f15a.eps}}
\vspace{1cm}
\centerline{
\includegraphics[width=8cm,angle=-90]{f15b.eps}}
\caption{Top: The distribution of $\ep$ versus 
the energy flux in the 15-150 keV band.  Two GRBs which locate in 
$\sim$ 7 $\times$ 10$^{-8}$ erg cm$^{-2}$ are GRB 050815 and GRB 050925.  One GRB 
which has $\ep$ of $\sim$ 20 keV is GRB 060428B.  Bottom: The distribution 
of $\ep$ versus the 1-s peak energy flux in the 15-150 keV band.}
\label{fig:fluence_pflux_ep}
\end{figure}

\begin{figure}
\centerline{
\includegraphics[width=10cm,angle=-90]{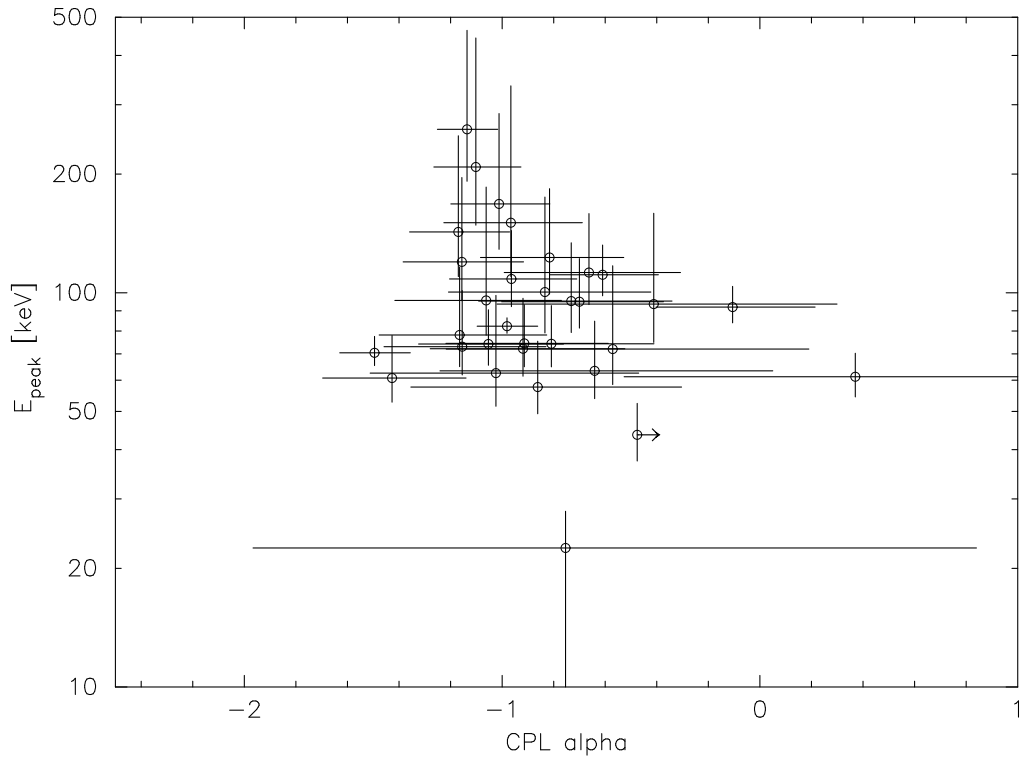}}
\caption{The distribution of $\eop$ versus the CPL photon index for 32
 GRBs which have a significant improvement in $\chi^{2}$ by a CPL fit over
 a PL fit.}
\label{fig:cpl_alpha_ep}
\end{figure}

For the limited number of GRBs (32 GRBs) which have a significant improvement 
in $\chi^{2}$ by a CPL fit, 
we investigated the relationship between $\eop$ and other parameters.  
The top panel of Figure \ref{fig:fluence_pflux_ep} shows the distribution of $\ep$ 
and the energy fluence in the 15-150 keV band.  The correlation coefficient 
between these two parameters is +0.580 (a chance probability of 0.2\%).  The 
bottom panel of Figure \ref{fig:fluence_pflux_ep} shows the distribution of $\ep$ 
and the 1-s peak energy flux in the 15-150 keV band.  The correlation 
coefficient is +0.563 (a chance probability of 0.2\%).  Note that the fluence 
and the peak flux are calculated from the BAT energy range.  Therefore, these 
values are not bolometric.  
Figure \ref{fig:cpl_alpha_ep} 
shows the relationship between the photon index in a CPL model and $\eop$.  
The $\eop$ distribution is spread from 10 keV to 500 keV for the BAT 
GRB sample.  There is no correlation between the photon index and $\eop$.  
This is consistent with the measurements of other missions 
\citep[e.g.,][]{kippen2001,sakamoto2005}.  

\begin{figure}
\centerline{
\includegraphics[width=10cm,angle=-90]{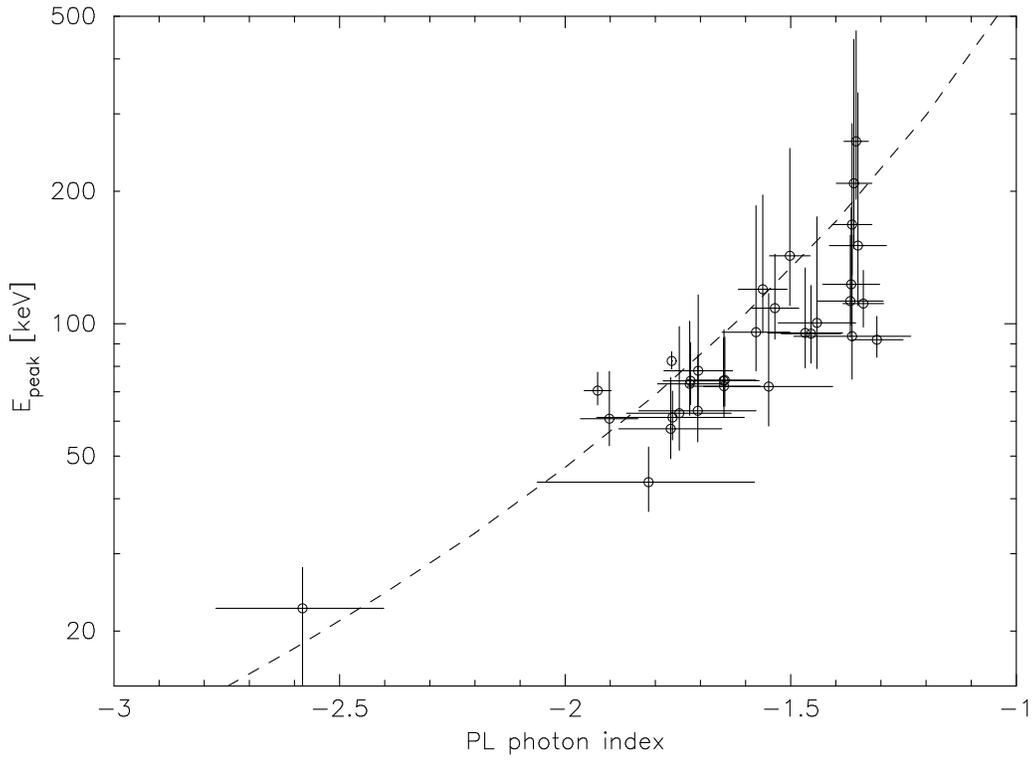}}
\caption{The distribution of $\eop$ versus the PL photon index for 32
 GRBs.  The dashed line is the PL photon index - $\eop$ correlation of 
\citet{zhang2007}: $\log \eop = 2.76 - 3.61 \log (-\alpha_{PL}$).}
\label{fig:ep_phindex}
\end{figure}

Figure \ref{fig:ep_phindex} shows the correlation between the photon index 
in a PL fit and $\eop$ derived from a CPL fit.  As we mentioned in the first 
paragraph of this section, the variation in the photon indices derived 
from a PL fit is very likely to reflect the differences in $\eop$ 
energies which are within the BAT energy band.  
This correlation between the PL photon 
index and $\eop$ was also recognized by \citet{zhang2007} and their best fit 
correlation is shown as a dashed line in Figure \ref{fig:ep_phindex}.  
Since 
the correlation found by \citet{zhang2007} is based on $\ep$ derived from a 
Band function, the slight offset between the dashed line and the data could be due 
to the systematic difference of $\ep$ based on the choise of the spectral model 
\citep[e.g.,][]{kaneko2006}.  
The detailed study of this correlation based on the spectral simulations will be 
presented elsewhere \citep{sakamoto2007}.  

\section{Summary}

The first BAT catalog includes 237 GRBs detected by BAT during two and a half 
years of operation.  
The BAT ground positions are $<$ 1.75$^{\prime}$ from the XRT position 
for 90\% of GRBs.  We presented the observed temporal and spectral properties 
of the prompt emission based on the analysis of the BAT event data.  Taking into 
account the difficulty in triggering short GRBs with the imaging instrument 
like BAT, the duration distributions are consistent with other missions.  
We showed that the BAT long GRB sample is systematically softer than that of 
the BATSE bright GRB sample.  The correlations such as the fluence- hardness 
and the peak flux - hardness have been confirmed by the BAT GRB samples.  

\acknowledgements
We would like to thank J.~A. Nousek, J.~P. Osborne, G. Chincarini, Y. Kaneko,  
and D. Malesani for valuable comments.  We also would like 
to thank the anonymous referee for comments and suggestions that materially
improved the paper.
This research was performed while T.~S. held a NASA Postdoctoral Program
administered by Oak Ridge Associated Universities at NASA Goddard Space
Flight Center formerly the National Research Council program.

\clearpage

\begin{deluxetable}{lccccccccccc}
\tabletypesize{\scriptsize}
\rotate
\tablecaption{BAT GRB summary\label{tbl:bat_summary}}
\tablewidth{0pt}
\tablehead{
\colhead{GRB} &
\colhead{Trigger} &
\colhead{Trigger time} &
\colhead{R.A.} &
\colhead{Dec.} &
\colhead{SN$_{img}$} &
\colhead{Error} & 
\colhead{T$_{90}$} &
\colhead{T$_{50}$} &
\colhead{Start} &
\colhead{Stop} &
\colhead{Comment}\\
\colhead{Name} &
\colhead{Number} &
\colhead{} &
\colhead{($^{\circ}$)} &
\colhead{($^{\circ}$)} &
\colhead{($\sigma$)} &
\colhead{($^{\prime}$)} &
\colhead{(s)} &
\colhead{(s)} &
\colhead{(s)} &
\colhead{(s)} &
\colhead{}}
\startdata
041217 & 100116 & 2004-12-17 07:28:25.236 & 164.793 & $ -17.942 $ & 20.4 & 1.3 & 5.8 & 2.7 & $ -2.2 $ & $ 17.8 $ & \\
041219A & 100318 & 2004-12-19 01:42:18.000 & 6.154 & $ 62.847 $ & - & 7.0 & - & - & - & - & No data$^{1}$\\
041219B & 100367 & 2004-12-19 15:38:48.000 & 167.674 & $ -33.458 $ & - & 7.0 & - & - & - & - & No data$^{1}$\\
041219C & 100380 & 2004-12-19 20:30:34.000 & 343.926 & $ -76.744 $ & 34.2 & 1.0 & 4.8 & 1.4 & $ -3.0 $ & $ 17.1 $ & \\
041220 & 100433 & 2004-12-20 22:58:26.599 & 291.288 & $ 60.598 $ & 33.8 & 1.0 & 5.6 & 2.2 & $ -299.9 $ & $ 302.1 $ & \\
041223 & 100585 & 2004-12-23 14:06:17.956 & 100.204 & $ -37.068 $ & 87.5 & 1.0 & 109.1 & 29.3 & $ -299.3 $ & $ 302.8 $ & \\
041224 & 100703 & 2004-12-24 20:20:57.698 & 56.192 & $ -6.663 $ & 11.3 & 2.0 & 177.2 & 37.5 & $ -299.0 $ & $ 303.0 $ & \\
041226 & 100815 & 2004-12-26 20:34:18.976 & 79.689 & $ 73.329 $ & 6.2 & 3.0 & 89.7 & 52.8 & $ -299.3 $ & $ 302.7 $ & \\
\enddata
\tablenotetext{1}{The event data are not available during this very early phase of operations.}
\tablenotetext{2}{{\tt battblocks} failed because of the weak short nature of the burst.}
\tablenotetext{3}{The event data of the last part of the burst emission are not available due to the downlink problem.}
\tablenotetext{4}{The event data of the brightest part of the burst emission are not available due to the entering to the SAA.}
\tablenotetext{5}{The burst duration was longer than the period of the event data.}
\tablenotetext{6}{\citet{markwardt2005}}
\tablenotetext{7}{The time interval calculated by the BAT flight software was used to create the image.}
\end{deluxetable}

\begin{deluxetable}{lccccccccccc}
\tabletypesize{\scriptsize}
\tablecaption{BAT GRB energy fluence \label{tbl:bat_fluence}}
\tablewidth{0pt}
\tablehead{
\colhead{GRB} &
\colhead{Spectral} &
\colhead{S(15-25)} &
\colhead{S(25-50)} &
\colhead{S(50-100)} &
\colhead{S(100-150)} &
\colhead{S(15-150)} &
\colhead{Start} &
\colhead{Stop}\\
\colhead{Name} &
\colhead{Model} &
\multicolumn{5}{c}{(10$^{-8}$ ergs cm$^{-2}$)} &
\colhead{(s)} &
\colhead{(s)}}
\startdata
041217 & CPL & $ 30.6 \pm 3.4 $ & $ 73.2 \pm 3.4 $ & $ 109.0 \pm 5.6 $ & $ 64.4 \pm 8.3 $ & $ 277.0 \pm 12.3 $ & $ +0.8 $ & $ +8.2 $ \\
041219A & - & - & - & - & - & - & - & - \\
041219B & - & - & - & - & - & - & - & - \\
041219C & PL & $ 29.6 \pm 2.6 $ & $ 39.7 \pm 2.1 $ & $ 39.2 \pm 3.1 $ & $ 22.7 \pm 2.9 $ & $ 131.0 \pm 7.3 $ & $ +6.7 $ & $ +14.2 $ \\
041220 & PL & $ 6.2 \pm 0.7 $ & $ 10.2 \pm 0.7 $ & $ 12.9 \pm 1.2 $ & $ 9.1 \pm 1.3 $ & $ 38.3 \pm 2.8 $ & $ -0.2 $ & $ +6.8 $ \\
041223 & PL & $ 143.0 \pm 6.1 $ & $ 334.0 \pm 8.3 $ & $ 617.0 \pm 10.6 $ & $ 581.0 \pm 16.4 $ & $ 1670.0 \pm 27.9 $ & $ -10.5 $ & $ +145.9 $ \\
041224 & CPL & $ 134.0 \pm 12.2 $ & $ 261.0 \pm 10.7 $ & $ 316.0 \pm 14.6 $ & $ 162.0 \pm 21.5 $ & $ 873.0 \pm 34.8 $ & $ -109.6 $ & $ +116.3 $ \\
041226 & PL & $ 3.9 \pm 1.6 $ & $ 7.5 \pm 1.8 $ & $ 11.4 \pm 3.4 $ & $ 9.2 \pm 4.4 $ & $ 32.1 \pm 8.3 $ & $ -1.3 $ & $ +93.4 $ \\
\enddata
\end{deluxetable}

\begin{deluxetable}{lccccccccccc}
\tabletypesize{\scriptsize}
\tablecaption{BAT GRB 1-s peak photon flux \label{tbl:bat_phflux}}
\tablewidth{0pt}
\tablehead{
\colhead{GRB} &
\colhead{Spectral} &
\colhead{F$^{\rm p}_{\rm ph}$(15-25)} &
\colhead{F$^{\rm p}_{\rm ph}$(25-50)} &
\colhead{F$^{\rm p}_{\rm ph}$(50-100)} &
\colhead{F$^{\rm p}_{\rm ph}$(100-150)} &
\colhead{F$^{\rm p}_{\rm ph}$(15-150)} &
\colhead{Start} &
\colhead{Stop}\\
\colhead{Name} &
\colhead{Model} &
\multicolumn{5}{c}{(photons cm$^{-2}$ s$^{-1}$)} &
\colhead{(s)} &
\colhead{(s)}}
\startdata
041217 & PL & $2.11 \pm 0.28$ & $2.25 \pm 0.18$ & $1.71 \pm 0.14$ & $0.80 \pm 0.11$ & $6.86 \pm 0.52$ & $ +3.3 $ & $ +4.3 $ \\
041219A & - & - & - & - & - & - & - & - \\
041219B & - & - & - & - & - & - & - & - \\
041219C & PL & $0.96 \pm 0.15$ & $0.82 \pm 0.09$ & $0.48 \pm 0.06$ & $0.18 \pm 0.04$ & $2.45 \pm 0.25$ & $ +8.1 $ & $ +9.1 $ \\
041220 & PL & $0.52 \pm 0.04$ & $0.58 \pm 0.03$ & $0.47 \pm 0.02$ & $0.23 \pm 0.02$ & $1.81 \pm 0.14$ & $ 0.0 $ & $ +1.0 $ \\
041223 & PL & $1.45 \pm 0.14$ & $2.12 \pm 0.12$ & $2.32 \pm 0.10$ & $1.45 \pm 0.10$ & $7.35 \pm 0.32$ & $ +35.0 $ & $ +36.0 $ \\
041224 & PL & $0.80 \pm 0.15$ & $0.94 \pm 0.11$ & $0.79 \pm 0.04$ & $0.41 \pm 0.03$ & $2.94 \pm 0.30$ & $ -0.1 $ & $ +0.9 $ \\
041226 & PL & $0.09 \pm 0.02$ & $0.11 \pm 0.02$ & $0.10 \pm 0.02$ & $0.05 \pm 0.01$ & $0.35 \pm 0.05$ & $ +3.3 $ & $ +4.3 $ \\
\enddata
\tablenotetext{\dagger}{The signal-to-noise ratio of the peak spectrum
 is too low to perform a spectral fit.}
\end{deluxetable}

\begin{deluxetable}{lccccccccccc}
\tabletypesize{\scriptsize}
\tablecaption{BAT GRB 1-s peak energy flux \label{tbl:bat_peakeneflux}}
\tablewidth{0pt}
\tablehead{
\colhead{GRB} &
\colhead{Spectral} &
\colhead{F$^{\rm p}_{\rm ene}$(15-25)} &
\colhead{F$^{\rm p}_{\rm ene}$(25-50)} &
\colhead{F$^{\rm p}_{\rm ene}$(50-100)} &
\colhead{F$^{\rm p}_{\rm ene}$(100-150)} &
\colhead{F$^{\rm p}_{\rm ene}$(15-150)} &
\colhead{Start} &
\colhead{Stop}\\
\colhead{Name} &
\colhead{Model} &
\multicolumn{5}{c}{(10$^{-8}$ ergs cm$^{-2}$ s$^{-1}$)} &
\colhead{(s)} &
\colhead{(s)}}
\startdata
041217 & PL & $6.5 \pm 0.8$ & $12.8 \pm 1.0$ & $19.4 \pm 1.7$ & $15.7 \pm 2.1$ & $54.4 \pm 4.2$ & $ +3.3 $ & $ +4.3 $ \\
041219A & - & - & - & - & - & - & - & - \\
041219B & - & - & - & - & - & - & - & - \\
041219C & PL & $2.9 \pm 0.5$ & $4.6 \pm 0.5$ & $5.4 \pm 0.7$ & $3.6 \pm 0.7$ & $16.6 \pm 1.8$ & $ +1.5 $ & $ +2.5 $ \\
041220 & PL & $1.6 \pm 0.2$ & $3.3 \pm 0.3$ & $5.4 \pm 0.5$ & $4.6 \pm 0.7$ & $15.0 \pm 1.3$ & $ -0.0 $ & $ +1.0 $ \\
041223 & PL & $4.6 \pm 0.4$ & $12.3 \pm 0.7$ & $26.9 \pm 1.2$ & $28.8 \pm 2.1$ & $72.6 \pm 3.3$ & $ +35.0 $ & $ +36.0 $ \\
041224 & PL & $2.5 \pm 0.5$ & $5.4 \pm 0.6$ & $9.1 \pm 1.0$ & $8.0 \pm 1.4$ & $25.0 \pm 2.6$ & $ -0.1 $ & $ +0.9 $ \\
041226 & PL & $0.3 \pm 0.1$ & $0.6 \pm 0.2$ & $1.1 \pm 0.4$ & $1.1 \pm 0.5$ & $3.1 \pm 0.9$ & $ +3.3 $ & $ +4.2 $ \\
\enddata
\tablenotetext{\dagger}{The signal-to-noise ratio of the peak spectrum
 is too low to perform a spectral fit.}
\end{deluxetable}

\begin{deluxetable}{lccc|cccc}
\tabletypesize{\scriptsize}
\tablecaption{BAT time-averaged spectral parameters \label{tbl:bat_spec}}
\tablewidth{0pt}
\tablehead{
\colhead{GRB} &
\colhead{$\alpha^{\rm PL}$} &
\colhead{K$^{\rm PL}_{50}$$^{\dagger}$} &
\colhead{$\chi^{2}$} & 
\colhead{$\alpha^{\rm CPL}$} & 
\colhead{$\eop$} &
\colhead{K$^{\rm CPL}_{50}$$^{\dagger}$} &
\colhead{$\chi^{2}$}
}
\startdata
041217 & $-1.46 \pm 0.07$ & $407.0 \pm 15.2$ & 74.9 & $-0.7 \pm 0.3$ & $ 95_{-14}^{+27}$ & $907_{-243}^{+363}$ & 54.8 \\
041219A & - & - & - & - & - & - & - \\
041219B & - & - & - & - & - & - & - \\
041219C & $-2.02 \pm 0.10$ & $70.9 \pm 4.1$ & 53.4 & - & - & - & - \\
041220 & $-1.66 \pm 0.12$ & $58.8 \pm 4.0$ & 28.9 & - & - & - & - \\
041223 & $-1.11 \pm 0.03$ & $103.0 \pm 1.9$ & 24.3 & - & - & - & - \\
041224 & $-1.72 \pm 0.06$ & $43.8 \pm 1.5$ & 56.1 & $-1.1 \pm 0.3$ & $ 74_{-9}^{+16}$ & $92_{-23}^{+34}$ & 36.7 \\
041226 & $-1.40 \pm 0.43$ & $3.5 \pm 0.8$ & 85.1 & - & - & - & - \\
\enddata
\tablenotetext{\dagger}{in 10$^{-4}$ photons cm$^{-2}$ s$^{-1}$ keV$^{-1}$.}
\end{deluxetable}

\newpage

\newpage

\newpage

\newpage

\newpage

\newpage

\newpage

\newpage

\newpage

\newpage

\newpage

\newpage

\newpage

\newpage

\newpage

\newpage

\newpage

\end{document}